\documentclass[aps,a4paper, preprint, superscriptaddress,preprintnumbers,floatfix,nofootinbib,amsmath,amssymb]{revtex4-1}

\usepackage{url}
\usepackage{hyperref}
\usepackage{color}
\usepackage{nameref}
\usepackage{cancel}
\usepackage{cleveref}
\usepackage{subfig}
\usepackage{soul}
\usepackage{colortbl}
\usepackage[table]{xcolor}

\usepackage{lipsum}
\usepackage[normalem]{ulem}

\usepackage{amstext,amssymb}
\usepackage{amsmath}
\usepackage{graphicx}
\usepackage{diagbox}
\usepackage{slashbox}
\usepackage{slashbox}
\usepackage{xspace}
\usepackage{color}
\usepackage{units}
\usepackage[T1]{fontenc}
\usepackage{amsmath,bm}
\usepackage{nicefrac}
\usepackage{slashed} 
\usepackage{multirow}
\newcommand{\be}{\begin{equation}}
\newcommand{\ee}{\end{equation}}
\newcommand{\bea}{\begin{eqnarray}}
\newcommand{\eea}{\end{eqnarray}}

\usepackage{slashed}

\newcommand{\nua}[1]{\ensuremath{\rlap{\kern-2.5pt\ensuremath{\overset{\scriptscriptstyle(-)}{\phantom{\nu}}}}{\ensuremath{{\nu}_{#1}}}}\xspace}

\newcommand{\deltaCP}{\ensuremath{\delta_{\rm CP}}}

\definecolor{brickred}{rgb}{0.8, 0.25, 0.33}
\definecolor{brightcerulean}{rgb}{0.11, 0.67, 0.84}
\definecolor{brown(traditional)}{rgb}{0.59, 0.29, 0.0}
\renewcommand{\thesection}{\arabic{section}}

\renewcommand{\thesubsubsection}{\thesection.\arabic{subsection}.\arabic{subsubsection}}
\usepackage{titlesec}

\titleformat{\subsubsection}[runin]{\normalfont\bfseries}{\thesubsubsection}{1em}{}

\begin{document}
\title{Study of large extra dimension and neutrino decay at P2SO experiment}
\author{Papia Panda}
\email{ppapia93@gmail.com}
\affiliation{School of Physics,  University of Hyderabad, Hyderabad - 500046,  India}
\author{Priya Mishra}
\email{mishpriya99@gmail.com}
\affiliation{School of Physics,  University of Hyderabad, Hyderabad - 500046,  India}
\author{Samiran Roy}
\email{samiranroy.hri@gmail.com}
\affiliation{School of Physics,  University of Hyderabad, Hyderabad - 500046,  India}
\author{Monojit Ghosh}
\email{mghosh@irb.hr}
\affiliation{Center of Excellence for Advanced Materials and Sensing Devices,
Ruđer Bošković Institute, 10000 Zagreb, Croatia}
\author{Rukmani Mohanta}
\email{rmsp@uohyd.ac.in}
\affiliation{School of Physics,  University of Hyderabad, Hyderabad - 500046,  India}
\begin{abstract}

In this study, we explore two intriguing new physics scenarios: the theory of Large Extra Dimensions (LED) and the theory of neutrino decay. We analyze the impact of LED on neutrino oscillations in the contexts of Protvino to Super-ORCA (P2SO), DUNE, and T2HK, with a particular emphasis on  P2SO. In contrast, the effects of neutrino decay are examined exclusively in the context of P2SO. For the LED scenario, we find that combining data from P2SO, DUNE, and T2HK can yield tighter constraints than current bounds, but only if all oscillation parameters are measured with high precision. In the case of neutrino decay, P2SO can achieve slightly better bounds compared to ESSnuSB and MOMENT, although its bounds remain weaker than those provided by DUNE and T2HK. Regarding sensitivities to unresolved oscillation parameters, the existence of LED has a minimal impact on the determination of CP violation, mass ordering and octant. However, neutrino decay can significantly influence the sensitivities related to CP violation and octant in a non-trivial manner.

\end{abstract}

\maketitle

\section{Introduction}
\label{intro}
The existence of neutrino mass has been conclusively established through the observation of neutrino flavor oscillations. The flavor and mass eigenstates of neutrinos are not identical; instead, they are related by the unitary Pontecorvo-Maki-Nakagawa-Sakata (PMNS) matrix ($U$), which is characterized by three mixing angles and one CP-violating phase. Neutrino oscillation physics has now entered an era of precision, with current and upcoming long-baseline experiments set to determine oscillation parameters with percent-level accuracy. This precision enables the investigation of sub-leading effects arising from various beyond the Standard Model (BSM) scenarios at neutrino detectors. In this paper, we focus on two such BSM scenarios: a) large extra dimensions (LED)~\cite{Arkani-Hamed:1998wuz,Dienes:1998sb,Dvali:1999cn,Barbieri:2000mg,Nortier:2021six} and b) neutrino decay \cite{Gelmini:1980re, DAmbrosio:1985aax, Oberauer:1987mg, Raghavan:1987uh, Gonzalez-Garcia:1988okv, Berezhiani:1991vk} in the context of the proposed long-baseline neutrino experiments, mainly focusing on Protvino to Super-ORCA (P2SO).
The first part of the paper focuses on LED, while the second part discusses the neutrino decay. 

 LED was proposed to address the hierarchy problem, which arises from the large discrepancy between the electroweak scale ($M_{\rm{EW}} \sim 10^3~ \rm{GeV}$) and the Planck scale ($M_{\rm{Pl}} \sim 10^{19} ~\rm{GeV}$), where gravitational effects become significant. This model assumes that there is only one fundamental scale, the electroweak scale. In four dimensions, the Planck scale is much larger than the electroweak scale, but in higher-dimensional space ($4+N$ dimensions), they become equivalent, \textit{i.e.}, $M_{\rm{Pl}} \sim M_{\rm{EW}}$.
In this framework, all Standard Model (SM) particles are confined to the four-dimensional space, while gravity can propagate into the higher dimensions. This propagation makes gravity appear much weaker in the four-dimensional space. The inclusion of higher dimensions also affects the known laws of gravity. For $N=1$, gravity would be modified at the solar system scale, and this scenario is ruled out by experimental observations. However, the $N=2$ case is consistent with experimental data.
We consider an asymmetric space where only one extra dimension is significantly larger than the others, effectively making the space five-dimensional.
Similar to gravity, the small neutrino mass can be naturally explained in this model. The right-handed SM singlet neutrino fields can propagate in the higher dimensions, 
and the suppression of the field in the 4-dimension by the volume of the extra dimension makes the neutrino mass very small~\cite{Arkani-Hamed:1998jmv,Antoniadis:1998ig,Arkani-Hamed:1998wuz}. When viewed from the 4-dimensional perspective, these singlet fields can be represented as a tower of Kaluza-Klein (KK) modes. These modes do not completely decouple from the system and exhibit mixing with the active neutrinos. Consequently, this mixing affects neutrino oscillations, providing a means to test the model in neutrino oscillation experiments. In recent years, numerous studies have been conducted to constrain the parameters of LED using neutrino oscillation experiments \cite{Machado:2011jt,MINOS:2016vvv,Forero:2022skg,Esmaili:2014esa,Basto-Gonzalez:2012nel,Basto-Gonzalez:2021aus,Rodejohann:2014eka,Carena:2017qhd,Stenico:2018jpl,Khan:2022bcl,Roy:2023dyq,Siyeon:2024pte,Giarnetti:2024mdt}. In this article, we investigate LED in the contexts of P2SO, Deep Underground Neutrino Experiment (DUNE), and Tokai-to-Hyper-Kamiokande (T2HK), with particular emphasis on P2SO. While studies of LED in the context of DUNE \cite{Giarnetti:2024mdt,Siyeon:2024pte,Calatayud-Cadenillas:2024wdw} and T2HK \cite{Roy:2023dyq} have been carried out previously, those works primarily focus on deriving bounds on the LED parameters. Moreover, Ref.~\cite{Forero:2022skg} demonstrates that the current bounds on LED parameters from MINOS/MINOS+, Daya Bay, and KATRIN are stronger than the projected bounds from DUNE and T2HK. The primary objective of this paper is to estimate the bounds on the LED parameter from P2SO and assess whether the combined constraints from DUNE, P2SO, and T2HK can improve upon the existing limits. We also explore the impact of marginalization over oscillation parameters and the effects of systematic uncertainties on these bounds. In addition, if LED exists in nature, it is important to explore how it could affect the CP violation, mass ordering, and octant sensitivities of these experiments. 
For the first time, we systematically examine the impact of LED on these standard sensitivities of P2SO, DUNE, and T2HK.

In the second part of the paper, we shift our focus on neutrino decay. The massive nature of neutrinos allows for the possibility of fast neutrino decay within BSM scenarios. Neutrinos could decay into either a lighter active neutrino or a sterile neutrino. When the final state includes an active neutrino, it is termed visible decay \cite{Abdullahi:2020rge,KamLAND:2003gfh,Pagliaroli:2015rca,Gago:2017zzy, Moss:2017pur, Coloma:2017zpg, Ascencio-Sosa:2018lbk,MacDonald:2024vtw}, whereas decay into a lighter sterile state is referred to as invisible decay \cite{Denton:2018aml,Martinez-Mirave:2024hfd}. This paper focuses on the invisible decay scenario. Depending on the nature of the neutrino \textit{i.e.}, whether it is a Dirac or Majorana particle, decay can occur through two different ways. For Dirac neutrinos, decay can produce a right-handed sterile neutrino along with an iso-singlet scalar~\cite{Acker:1991ej,Acker:1992eh}. In the case of Majorana neutrinos, decay can result in a sterile neutrino and a Majoron\cite{Chikashige:1980ui, Schechter:1981cv, Gelmini:1980re, Gelmini:1983ea}. The effect of decay on neutrino oscillation depends on the mass ($m_i$) and lifetime ($\tau_i$) of the neutrino, represented by the factor exp$\left(-\dfrac{m_i \, L}{\tau_i \, E_{\nu}}\right)$, where $L$ and $E_{\nu}$ correspond to the length of propagation and energy of the neutrino, respectively. Various experiments place strong constraints on the neutrino decay parameter ($\tau_i/m_i$). Solar neutrino data establishes a lower bound on $\tau_2/m_2$ 
\cite{Bahcall:1972my,Berryman:2014qha, Picoreti:2015ika, Huang:2018nxj, Bandyopadhyay:2002qg}, assuming the decay of the $\nu_2$  state only. The KamLAND reactor experiment constrains the decay parameter $\tau_1/m_1$ \cite{Berryman:2014qha}. Additionally, supernova neutrino data sets stringent constraints on the neutrino decay hypothesis. Observations of neutrinos from supernova SN1987A place strong constraint on the decay of $\nu_e$ state \cite{Frieman:1987as}.  
Analysis of MINOS/MINOS+, T2K and NO$\nu$A data provided bounds on $\tau_3/m_3$ \cite{Gomes:2014yua,Choubey:2018cfz,Ternes:2024qui}. 
The differences in track and cascade spectra of the IceCube data prefers the invisible neutrino decay scenario at more than $3\sigma$ C.L.~\cite{Denton:2018aml,Abdullahi:2020rge}. 
Projected sensitivities from the upcoming experiments such as DUNE, ESSnuSB, and T2HK on $\tau_3/m_3$ can be found in \cite{Choubey:2017dyu, Ghoshal:2020hyo,Choubey:2020dhw, Chakraborty:2020cfu, Dey:2024nzm}. These bounds are stronger than the current bounds on $\tau_3/m_3$. In this paper, for the first time we examine the effect of decay in the P2SO experiment and provide the projected bounds on $\tau_3/m_3$ parameter, and compare our results with existing constraints. We also show how the marginalization of oscillation parameters affects $\tau_3/m_3$ bounds and the effect of decay in measuring CP violation and octant sensitivities.

The structure of this paper is as follows: Section~\ref{led-exp} presents an overview of the key features of the three long-baseline neutrino experiments P2SO, DUNE, and T2HK. Section~\ref{Simulation-details} outlines the statistical methods and simulation techniques employed in this work. The analysis of the LED scenario is detailed in Section~\ref{led}, while Section~\ref{sec:decay} focuses on the study of invisible neutrino decay. Finally, Section~\ref{conclu} summarizes our findings and concludes the paper, highlighting the results from both new physics scenarios.


\section{Experimental details}
\label{led-exp}

In this section, we outline the key details of three experiments considered in the analysis for the LED and neutrino decay.   

\subsection{P2SO}
\label{led-p2so}
In the case of the P2SO experiment, which is a future long-baseline neutrino experiment with a baseline of 2595 km, we formulate our simulation details using the technical design report from Refs.~\cite{Akindinov:2019flp, Hofestadt:2019whx}. For more detailed description of the P2SO experimental setup, see the Refs.~\cite{Singha:2022btw, Majhi:2022fed, Singha:2023set}. The experiment will feature a few megatonnes of fiducial detector volume and a beam power of 450 kW, corresponding to $4 \times 10^{20}$ protons-on-target (POT). For the purpose of simulation, we consider a total runtime of 6 years, with 3 years dedicated to neutrino mode and 3 years to antineutrino mode. The systematic uncertainty values are taken from Ref.~\cite{Akindinov:2019flp}.

\subsection{DUNE}
\label{led-dune}
The DUNE is one of the most promising upcoming long-baseline neutrino experiments, with a 1300 km baseline spanning from the Fermi National Accelerator Laboratory (FNAL) to the Sanford Underground Research Facility (SURF). For the simulation of DUNE experiment, we use the official files associated with the technical design report \cite{DUNE:2021cuw}. The files represent an exposure of 624 kt-MW-years which corresponds to 6.5 years of run each in neutrino (FHC) and antineutrino (RHC) modes, using a 40 kt fiducial mass liquid argon time-projection chamber (LArTPC) far detector and a 120-GeV, 1.2 MW beam. This configuration is equivalent to ten years of data collection, following the nominal staging assumptions outlined in \cite{DUNE:2020jqi}. For systematic errors, we use the numerical values from Ref.~ \cite{DUNE:2021cuw}.

\subsection{T2HK}
\label{led-t2hk}
The T2HK experiment is another proposed long-baseline neutrino project, with a 295 km baseline and an off-axis angle of $2.5^\circ$, producing a very narrow neutrino beam. For the T2HK simulation, we adopt the configuration details from Ref.~\cite{Hyper-Kamiokande:2016srs}. The neutrino source, located at J-PARC, will operate with a beam power of 1.3 MW, delivering a total exposure of $27 \times 10^{21}$ protons-on-target (POT), which is equivalent to ten years of operation. We consider an equal runtime for neutrino and anti-neutrino modes; each of five years. For the systematic errors, we take the values from the paper \cite{Hyper-Kamiokande:2016srs}. The detector technology will utilize a water Cherenkov detector with a fiducial volume of 374 kt. 

\section{Statistical method and simulation details}
\label{Simulation-details}
To simulate P2SO, DUNE and T2HK experiments, we use the General Long-Baseline Experiment Simulator (GLoBES) software \cite{Huber:2004ka, Huber:2007ji}. We have modified the probability engine to incorporate the effects of the large extra dimension scenario and neutrino decay. This engine calculates the exact neutrino oscillation probabilities in matter. To estimate sensitivity we consider the Poisson log-likelihood formula:

\begin{equation}
    \chi^2 = 2 \sum_{i=1}^n \left[N_i^{\rm test} - N_i^{\rm true} - N_i^{\rm true} \log \left( \frac{N_i^{ \rm test}}{N_i^{\rm true}} \right) \right],
    \label{chi}
\end{equation}
where $N_{i}^{\rm true}$ and $N_{i}^{\rm test}$ represent the event numbers in the true and test spectra, respectively, and ``i'' denotes the number of energy bins. The true values of the oscillation parameters are taken from Table~ \ref{Tab:osc}. All the relevant oscillation parameters are marginalized in our analysis. We use the method of pull \cite{Gonzalez-Garcia:2004pka, Fogli:2002pt} to include the effect of systematic uncertainties. For systematic errors, we consider an overall normalization error corresponding to signal and background. We present our results considering normal ordering of the neutrino masses.
Throughout the simulation, we consider only charged current (CC) interactions as the signal, since neutral current (NC) interactions have a marginal impact on the bounds in both scenarios.

\label{led-theo}

 \begin{table}[]
    \centering
    \begin{tabular}{||c||c||}
    \hline
    \hline
      \hspace{1cm}Oscillation parameters\hspace{1cm}  & \hspace{1cm}Best-fit values $\pm$ $1\sigma$\hspace{1cm} \\
       \hline
       \hline
       $\sin^2 \theta_{12}$ &$0.303^{+ 0.012}_{- 0.012}$\\
        \hline
       $\sin^2 \theta_{13}$ & $0.02225^{+ 0.00056}_{- 0.00059}$ \\
        \hline
       $\sin^2 \theta_{23}$  &  $0.448^{+ 0.019}_{- 0.016}$   \\
        \hline
        $\delta_{CP}$  &   $270^{\circ}$ \\
        \hline
        $\Delta m_{21}^2  (\rm eV^2)$  & $7.41^{+ 0.21}_{- 0.20} \times 10^{-5}$ \\
        \hline
        $\Delta m_{31}^2 (\rm eV^2)$  & $ 2.507^{+ 0.026}_{- 0.027} \times 10^{-3}$\\
        \hline
        \hline
    \end{tabular}
    \caption{Values of the oscillation parameters used in our analysis, for both LED and neutrino decay, are taken from Ref.~\cite{Esteban:2020cvm}. We consider normal ordering for the entire analysis and vary $\deltaCP$ in full range.}
    \label{Tab:osc}
\end{table}
 
\section{Large Extra dimension}
\label{led}

In this section, we discuss the framework of LED, and in the subsequent section, we examine neutrino decay. We begin with a brief overview of the theoretical framework of LED in the context of neutrino oscillations, followed by an exploration of its impact on neutrino oscillation probabilities and the expected event rates.
Next, we estimate the bounds on LED parameters in the context of  P2SO, DUNE+T2HK, and DUNE+T2HK+P2SO experimental configurations. In this context, we also analyze the effects of systematic uncertainties and the impact of marginalization of the oscillation parameters.
Finally, for the very first time we show how the CP violation, mass ordering and octant sensitivities of  individual experiments i.e., P2SO, DUNE and T2HK get altered in presence of LED. We explained our numerical results using analytical expressions.

\subsection{Theoretical framework}
\label{theo}
In the framework of LED, all the SM particles are restricted to four dimensional space, while the gravity could propagate through all dimensions, including the large extra dimensions. This produces weak gravitational field in the four dimensional space. Similar to the gravity, we can generate the small neutrino mass by introducing the SM singlet neutrinos that propagate all dimensions. We extend the SM sector by adding three 5-D singlet fermionic fields $\Psi^\alpha_{L,R}$ corresponding to three SM active neutrino fields $\nu^{\alpha}_L$. After the compactification of the fifth dimension on a circle of radius $R_{\rm{ED}}$, those fields can be decomposed as a tower of Kaluza-Klein (KK) modes ($\psi^{\alpha (n)}_{L,R}\, , n=-\infty ..  \infty$). The fields that couple to the SM neutrinos are redefined as, $\nu^{\alpha (0)}_{R} \equiv \psi^{\alpha (0)}_{R}$ and $\nu^{\alpha(n)}_{L,R} \equiv (\psi^{\alpha (n)}_{L,R} + \psi^{\alpha (-n)}_{L,R} )/ \sqrt{2}$. 
Using this notation, the mass term of the Lagrangian~\cite{Davoudiasl:2002fq} can be expressed as
\begin{eqnarray} 
L_{\rm{mass}} = m^{D}_{\alpha \beta} \big(\bar{\nu}^{\alpha (0)}_{R} \nu^{\beta}_{L} + \sqrt{2} \sum^{\infty}_{n=1} \bar{\nu}^{\alpha (n)}_{R} \nu^{\beta}_{L} \big)
 +  \sum^{\infty}_{n=1} \dfrac{n}{R_{\rm{ED}}} \bar{\nu}^{\alpha(n)}_{R} \nu^{\alpha(n)}_{L} + h.c. \, ,
 \label{mass_matrix}
\end{eqnarray}
where $m^D$ is the Dirac mass matrix. 
The diagonalization of the mass matrix is carried out in two steps. We first introduce two $3\times 3$ matrices $U$ and $r$ that diagonalize $m^D$ \textit{i.e.} $m^D_{\rm{diag}} = r^{\dagger} m^D U=\rm{diag} (m^D_1, m^D_2, m^D_3)$ and  
\begin{eqnarray}
\nu^{\alpha}_L &=& U^{\alpha i} \nu^{\prime i(0)}_L \\
\nu^{\alpha (n)}_R &=& r^{\alpha i } \nu^{\prime i(n)}_R, n=0...\infty \\
\nu^{\alpha (n)}_L &=& r^{\alpha i} \nu^{\prime i(n)}_L, n=1...\infty .
\end{eqnarray}
In the pseudo mass basis,  $\nu^{\prime i}_{L}=\big (\nu^{\prime i},\nu^{\prime i(1)},\nu^{\prime i(2)}, .. \big )^T_{L}  $ and $\nu^{\prime i}_{R}=\big (\nu^{\prime i(0)},\nu^{\prime i(1)},\nu^{\prime i(2)}, .. \big )^T_{R}  $, the mass term in Eq.~\ref{mass_matrix} takes the form 
\begin{eqnarray}
L_{\rm{mass}} =\sum^3_{i=1} \bar{\nu}^{\prime i}_R M^i \nu^{\prime i}_L+ h.c.
 \label{mass_matrix1}
\end{eqnarray}
where $M_i$ represents an infinite-dimensional matrix, 
\begin{eqnarray}
M^i= \dfrac{1}{R_{\rm{ED}}}
\begin{pmatrix}
m_i^DR_{\rm{ED}}&0&0&0&\ldots\\
\sqrt{2}m_i^D R_{\rm{ED}}&1&0&0&\ldots\\
\sqrt{2}m_i^DR_{\rm{ED}}&0&2&0&\ldots\\
\vdots&\vdots&\vdots&\vdots&\ddots
\end{pmatrix}.
\end{eqnarray}
The infinite-dimensional matrix $M^i$ can be diagonalized to obtain the true mass basis.
We need two infinite-dimensional matrices ($L$ and $R$) for the diagonalization of $M^i$ which makes $R^{\dagger}_i M^i L_i$, a diagonal matrix. The actual mass basis is related to the pseudo mass basis by $\nu^i_L = L^{\dagger} \nu^{\prime i}_L $ and $\nu^i_R = R^{\dagger} \nu^{\prime i}_R $. The flavor neutrinos at the four dimensional space are related to the mass basis as
\begin{eqnarray}
\nu^{\alpha}_L = \sum^{3}_{i=1} U^{\alpha i} \sum^{\infty}_{n=0} L^{0n}_i \nu^{i(n)}_L.
\end{eqnarray}
Here, $L$ can be calculated by diagonalizing the Hermitian matrix $M^{\dagger} M$~\cite{Arkani-Hamed:1998wuz,Dienes:1998sb,Dvali:1999cn,Barbieri:2000mg} as
\begin{eqnarray}
\left(L^{0n}_i\right)^2=\frac{2}{1+\pi^2\left(m_i^{D}R_{\rm{ED}}\right)^2+\left(\lambda^{(n)}_i\right)^2/\left(m_i^{D}R_{\rm{ED}}\right)^2}.
\end{eqnarray}
The eigenvalues of the matrices $R^2_{\rm{ED}} M^{\dagger}_i M_i$ are represented by $\left(\lambda^{(n)}_i\right)^2$. These values can be obtained by solving the following equation
\begin{equation}\label{trans}
\lambda_i^{(n)}-\pi \big(m_i^DR_{\rm{ED}}\big)^2\cot\left(\pi\lambda_i^{(n)}\right)=0.
\end{equation}
The mass of $\nu^{i(n)}_L$ is $\lambda^{(n)}_i/R_{\rm{ED}}$ and 
\begin{eqnarray}
L^{jn}_i = \dfrac{\sqrt{2} j m^D_i R_{\rm{ED}}}{ (\lambda^{(n)}_i)^2-j^2} L^{0n}_i,
\end{eqnarray}
where $j=1..\infty$ and $n=0..\infty$. 
We focus on the scenario where the impact of LED can be perceived as a small perturbation to the standard neutrino oscillation and this suggests that $m^D_i R_{\rm{ED}}<<1$ \footnote{This assumption is adopted throughout the analysis.}. On the basis of this assumption, we can write
\begin{eqnarray}\nonumber
\lambda^{(0)}_i = m^D_i R_{\rm{ED}} \big(1-\dfrac{\pi^2}{6}(m^D_i R_{\rm{ED}})^2+..\big), ~~~~~~~ \lambda^{(j)}_i = j+\dfrac{1}{j}(m^D_i R_{\rm{ED}})^2+..\\ \nonumber
L^{00}_i = 1-\dfrac{\pi^2}{6}(m^D_i R_{\rm{ED}})^2+..,~~~~~~~~~~~~~~~~~~~ L^{0j}_i= \dfrac{\sqrt{2} m^D_i R_{\rm{ED}}}{j} + .. ~~~~~~ \\
L^{j0}_i= -\dfrac{\sqrt{2} m^D_i R_{\rm{ED}}}{j} + .., ~~~~~~~~~~~~~~~~~~~~~~~~~~ L^{jj}_i=1- \dfrac{( m^D_i R_{\rm{ED}})^2}{j^2} + ..,~
\label{lambda_L}
\end{eqnarray}
and $L^{kj}=\mathcal{O}(( m^D_i R_{\rm{ED}})^2)$ for $k \neq j=1..\infty$.
In the presence of LED, the oscillation probability of a specific neutrino flavor $\nu_{\alpha}$ to $\nu_{\beta}$ is given by
\begin{eqnarray}
P_{\alpha\beta}(L,E_\nu) = \big |\sum^{3}_{i=1} U^{\alpha i} U^{*\beta i } A_i(L,E_\nu)\big |^2,
\label{CC_p}
\end{eqnarray}
where 
\begin{eqnarray}
A_i(L,E_\nu) = \sum^{\infty}_{n=0} \left(L^{0n}_i\right)^2 {\rm{exp}} \left(i \dfrac{\lambda_i^{(n)2} L}{2E_\nu R^2_{\rm{ED}}} \right).
\label{CC_p_e}
\end{eqnarray}
The first term of Eq.~\ref{lambda_L} relates Dirac masses ($m^D_i$) and the neutrino masses ($\lambda^{(0)}_i/R_{\rm{ED}}$) of the active neutrinos. From this, we can write $\Delta m^2_{ij} R^2_{\rm{ED}} = (\lambda^{(0)}_i)^2 - (\lambda^{(0)}_j)^2 $. Two parameters ($m^D_2, m^D_3$) can be eliminated from the theory using the known values of the solar ($\Delta m^2_{21}$) and atmospheric ($\Delta m^2_{31}$) mass squared differences. As a result, the oscillation probability depends on the two extra parameters, $m^D_1~(\equiv m_0)$ and $R_{\rm{ED}}$. 
Matter modifies the vacuum neutrino oscillation probability which in the presence of LED is governed by the following equation~\cite{Berryman:2016szd}:
\begin{eqnarray}\label{evolution}
i\frac{d}{dt}{\nu^{\prime}_i}_L=\Bigg[\frac{1}{2E_\nu}M_i^{\dagger}M_i{\nu^{\prime}_i}_L+\sum_{j=1}^3
\begin{pmatrix}
V_{ij} & 0_{1\times n} \\
0_{n\times 1} & 0_{n\times n}
\end{pmatrix}
{\nu^{\prime}_i}_L\Bigg]_{n\to\infty},~ V_{ij}=\sum_{\alpha=e,\mu,\tau} U^*_{\alpha i}U_{\alpha j}\Big(\delta_{\alpha e}V_{\rm{CC}}+V_{\rm{NC}}\Big),\nonumber\\
\end{eqnarray}
where the charged and neutral current matter potentials are represented by $V_{CC}=\sqrt{2}G_Fn_e$ and $V_{NC}=-1/\sqrt{2}\, G_Fn_n$ respectively. The electron and neutron number densities are denoted by $n_e$ and $n_n$, respectively. For various baselines, we keep the matter density constant during the neutrino evolution while taking into account the equal number density of electrons and neutrons. For our numerical analysis, we assume two KK modes, and we have checked that, the inclusion of the larger number of modes has minimal effect on the outcome.

\begin{figure}[htbp]
\includegraphics[height=60mm, width=81mm]{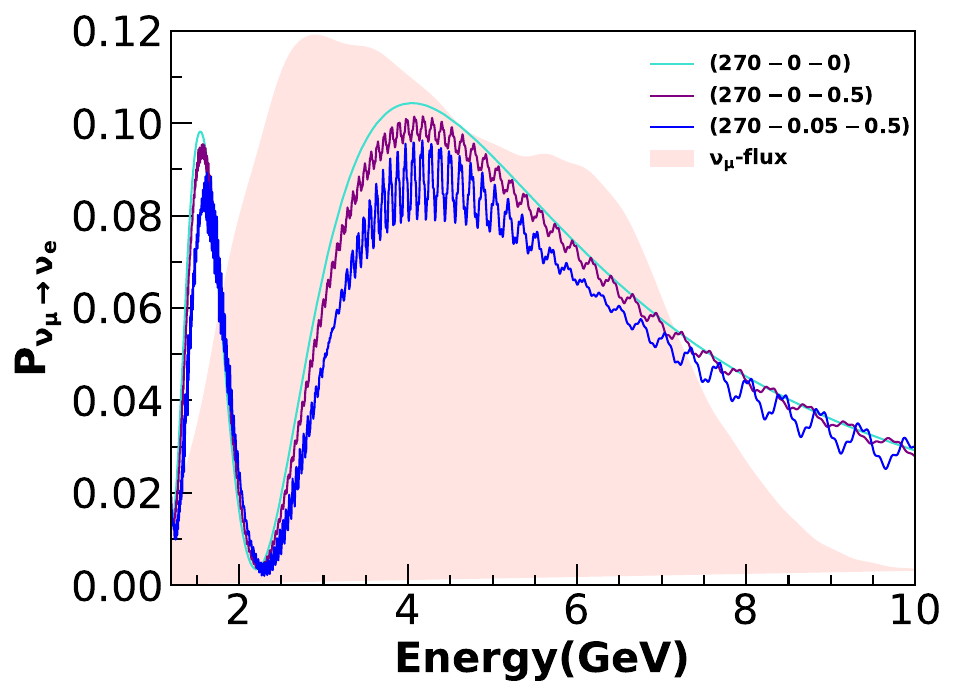}    \includegraphics[height=60mm, width=81mm]{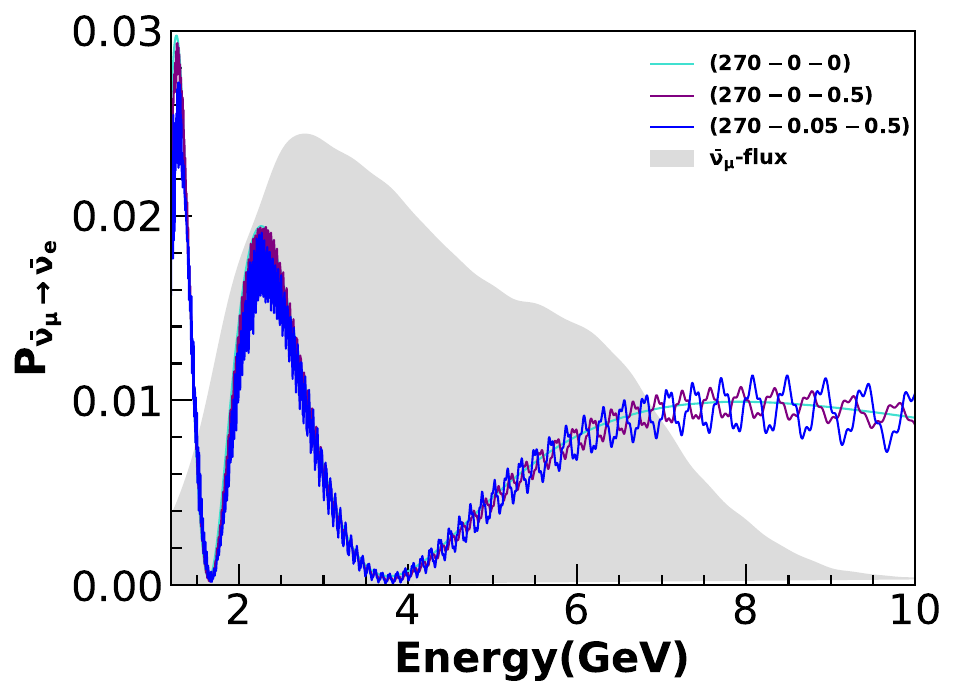}\\
  \includegraphics[height=60mm, width=81mm]{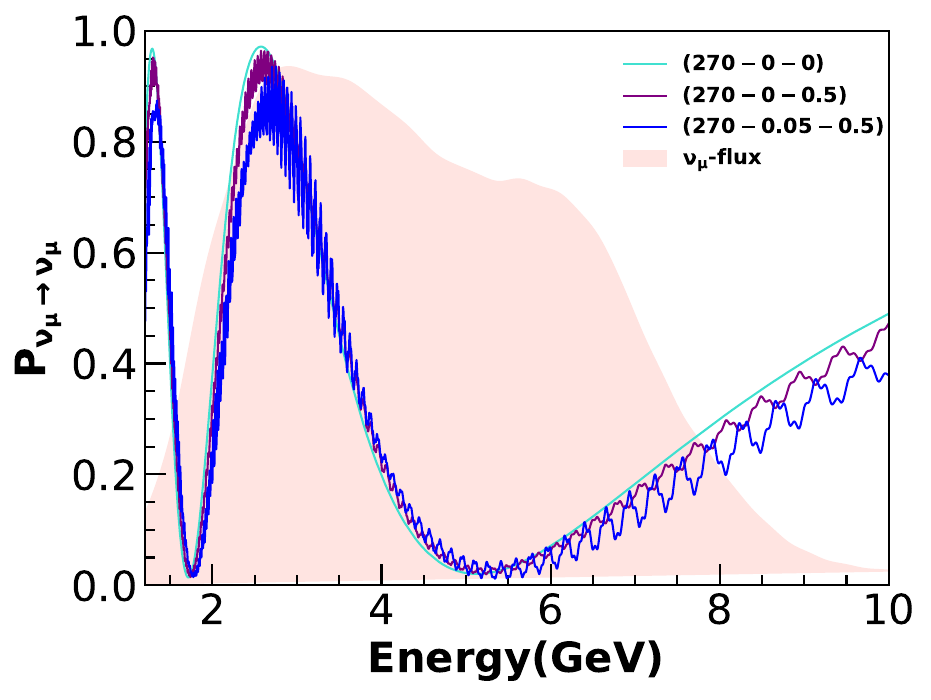}     \includegraphics[height=60mm, width=81mm]{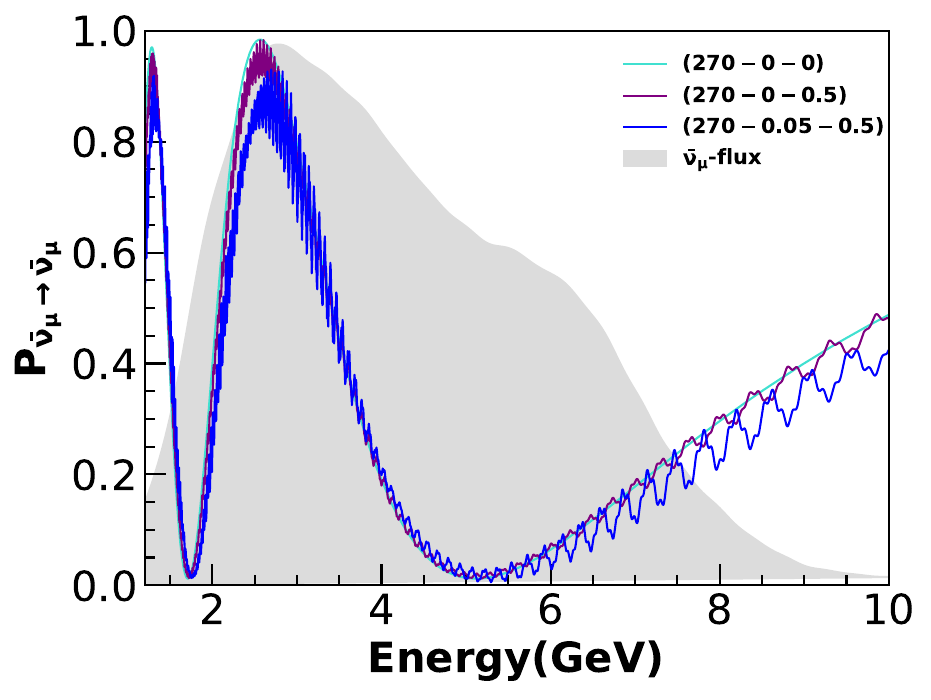}
     \caption{Probability plot as a function of neutrino energy for P2SO. Upper (lower) row shows the appearance (disappearance) probability for different combinations of $m_0$ and $R_{ED}$ values. Left (right) column shows the results for neutrino (antineutrino). The legend of each panel has the form $(\delta_{\rm CP}~ [^{\circ}]-m_0~[\rm eV]- R_{ED}~[\mu \rm m)]$.}
     \label{prob3}
 \end{figure}
 
\subsection{Probability and event rates in presence of LED}
\label{led-prob}

In this section, we discuss the behavior of the oscillation probabilities and event rates in presence of LED. Fig.~\ref{prob3} represents the appearance and disappearance probabilities for neutrino and antineutrino modes. The left column corresponds to neutrinos, and the right column to antineutrinos. The upper row displays appearance probabilities, while the lower row shows disappearance probabilities under various conditions. The light orange (gray) shaded region depicts the shape of the $\nu_{\mu} ~(\bar{\nu}_{\mu})$ flux of the P2SO experiment. For this figure we consider $\delta_{\rm CP} = 270^\circ$. The cyan curves depict the probabilities without LED, while brown (blue) curve shows the probability with $R_{ED} = 0.5~ \mu {\rm m}$ keeping $m_0=0~(0.05)$ eV. 

From the panels we see that the presence of LED parameters results a decrease of oscillation probability from standard case. Most importantly, here we also see a distortion in the spectrum due to fast oscillations. This distortion increases when we consider non-zero value of $m_0$.  To understand these behaviors, we calculate the analytical probability expression in vacuum for the electron neutrino appearance channel as \footnote{Note that though Eq. \ref{eq app} is derived in vacuum, this equation is sufficient to explain the main features of Fig.~\ref{prob3} which is generated in matter.},
 \begin{eqnarray}
     P_{\mu e} (L,E_{\nu}) &\simeq& P_{\mu e}^{\rm SI} (L,E_{\nu}) +  R_{ED}^2  \Big[A +  B \cos\Big(\frac{L \Delta m_{31}^2}{2 E_{\nu}} + \deltaCP\Big) + C \Big(\cos\Big(\frac{L \Delta m_{31}^2}{2 E_{\nu}} - \frac{L}{2 E_{\nu} R_{ED}^2}\Big) \nonumber \\ &-&  \cos\Big(\frac{L}{2 E_{\nu} R_{ED}^2}\Big) \Big)\Big] 
     \label{eq app},
 \end{eqnarray}

where,
\begin{eqnarray}
A&=& 1.2 \times 10^{-5} \cos \delta_{\rm CP} \sin 2 \theta_{23}    ~~\rm{eV}^2 \nonumber \\
B&=& -1.6 \times 10^{-5} \sin 2 \theta_{23} ~~\rm{eV}^2   \nonumber \\ 
C&=&0.0871  \Delta m_{31}^2 \sin^2 \theta_{23}.
\label{a-value}
\end{eqnarray}

In the above equation, we consider $m_0 = 0$ eV. Here, $R_{ED}$, $L$ and $E_{\nu}$ are in eV$^{-1}$, eV$^{-1}$ and eV respectively. The expression consists of two terms: the first, $P_{\mu e}^{\rm SI}$, represents the standard appearance probability \cite{Choubey:2003yp}, and the second term incorporates the dependence on $R_{ED}$. Here we note that the LED term is proportional to $R_{ED}^2$. By putting the values of the parameters A, B, and C, it can be shown that an overall negative sign appears with the new physics LED term. This explains why the probabilities in presence of LED are smaller than the standard scenario. In this equation we clearly identify the factor $L/2 E_{\nu} R_{ED}^2$ which is responsible for the fast oscillation. Additionally, a non-zero value of $m_0$ can also amplify these fast oscillations (cf. Eq.~\ref{CC_p_e}). This is true for both neutrinos and antineutrinos in the appearance and disappearance channels.

\begin{figure}[htbp]
     \includegraphics[scale=0.5]{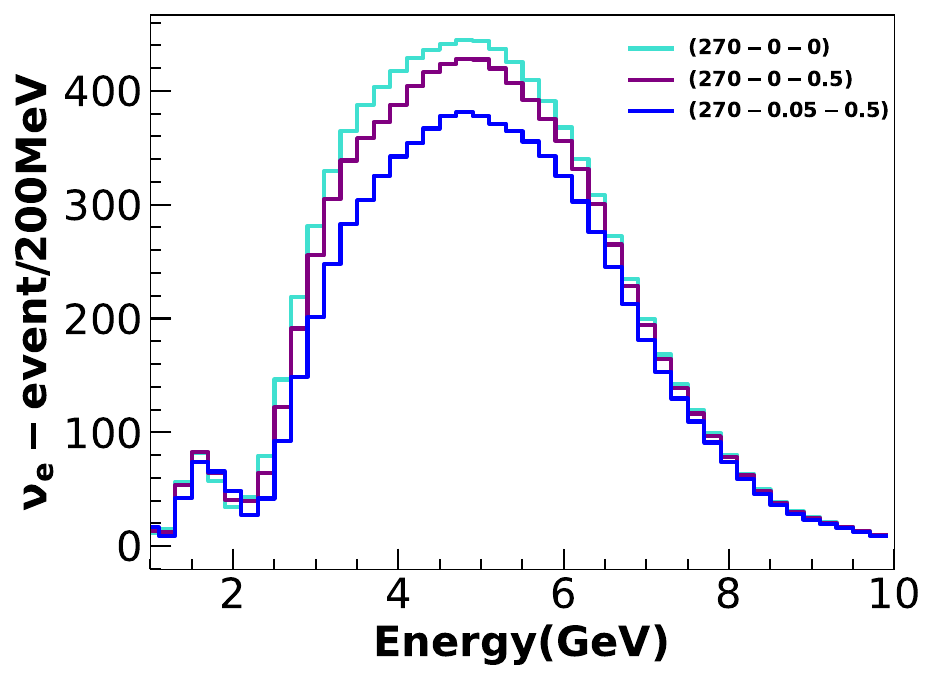}
     \includegraphics[scale=0.5]{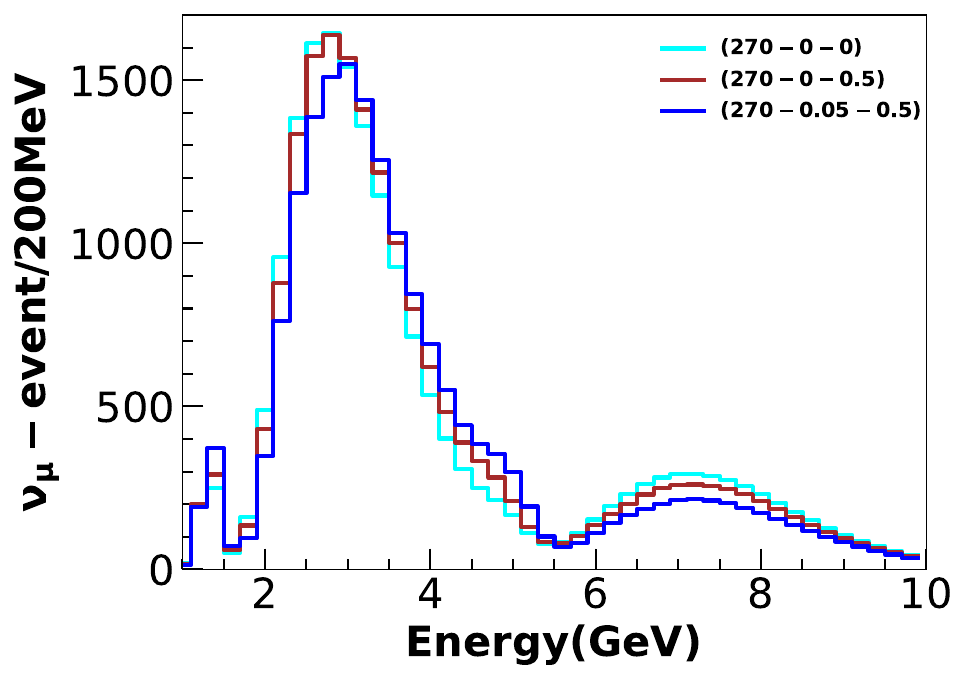}
     \caption{Event rates in the presence and absence of LED parameters for P2SO. Left (right) panel depicts the event rate for appearance (disappearance) channel. Color code is given in the legend. The legend of each panel has the form $(\delta_{\rm CP}~[^\circ]-m_0~[\rm eV]- R_{ED}~[\mu \rm m)]$.}
     \label{event2}
 \end{figure}

In Fig. \ref{event2}, we show the event rates for electron and muon neutrinos in the P2SO experiment, with and without the LED parameters using the same color scheme as in Fig. 
\ref{prob3}. 
The left (right) panel of the figure 
shows the event rates for the appearance (disappearance) channel under different LED parameter conditions. We observe that, with a non-zero value of $R_{ED}$, the event rate decreases, and the overall amplitude decreases even further when $m_0$ is non-zero. The nature of the event rate is identical to the probability plots shown in Fig. \ref{prob3}.
 
 \subsection{Results for LED}
\label{led-result}

\subsubsection{Bound on LED parameters \\}
\label{led-bound}

In this section, we project the bounds on the LED parameters $m_0$ and $R_{ED}$, in different combinations of future long-baseline neutrino experiments. As mentioned in the introduction, the constraints on LED parameters have been previously studied for DUNE~\cite{Roy:2023dyq} and T2HK~\cite{Roy:2023dyq}.
In this study, we aim to explore whether combining these experiments can impose stronger constraints than the P2SO experiment and identify the conditions under which this is possible.

\begin{table}
    \centering
    \begin{tabular}{||c||c||c||}
    \hline
    \hline
      Setup  & Conditions  &  $R_{ED}$($\mu m$) at $m_0$=0 eV\\
      \hline
      \hline
            P2SO
            &  all-fixed-no-sys  & 0.194 \\
              \hline
              & all-fixed-with-sys & 0.232   \\
            \hline
            &  $\delta_{\rm CP}$-free-with-sys & 0.230\\
                \hline
             &  $\delta_{\rm CP}-\theta_{23}-\Delta m_{31}^2$-free-no-sys  & 0.236 \\
              \hline 
                  &  $\delta_{\rm CP}-\theta_{23}$-free-with-sys  & 0.265 \\
                  \hline
                  &  $\Delta m_{31}^2$-free-with-sys  & 0.345 \\
                    \hline
                    &  $\delta_{\rm CP}-\theta_{23}-\Delta m_{31}^2$-free-with-sys  & 0.361 \\
                    \hline
                    & all-free-with-sys  & 0.361\\
                    \hline
              \hline
              DUNE+T2HK  
       &  all-fixed-no-sys  & 0.229 \\
       \hline 
        &  $\delta_{\rm CP}-\theta_{23}-\Delta m_{31}^2$-free-no-sys  & 0.235 \\
            \hline
      & all-fixed-with-sys & 0.317  \\
            \hline
                &  $\delta_{\rm CP}$-free-with-sys & 0.317 \\
                \hline
                  &  $\delta_{\rm CP}-\theta_{23}$-free-with-sys  & 0.317 \\
                  \hline
                  &  $\Delta m_{31}^2$-free-with-sys  & 0.390 \\
                    \hline
                    &  $\delta_{\rm{CP}}-\theta_{23}-\Delta m_{31}^2$-free-with-sys  & 0.414 \\
                    \hline
                    & all-free-with-sys  & 0.414\\
                    \hline
    \hline
DUNE+T2HK+P2SO  &  all-fixed-no-sys  & 0.175 \\
              \hline 
&  all-fixed-with-sys &  0.208   \\
            \hline
                &  $\delta_{\rm CP}$-free-with-sys & 0.215 \\
                \hline
                &  $\delta_{\rm CP}-\theta_{23}-\Delta m_{31}^2$-free-no-sys  & 0.222 \\
            \hline
                  &  $\delta_{\rm CP}-\theta_{23}$-free-with-sys  & 0.232 \\
                  \hline
                  &  $\Delta m_{31}^2$-free-with-sys  &  0.299 \\
                  \hline
                    &  $\delta_{\rm CP}-\theta_{23}-\Delta m_{31}^2$-free-with-sys  &  0.320 \\
                    \hline
                    & all-free-with-sys  &  0.320\\
                    \hline
            \hline
    \end{tabular}
    \caption{Bounds on $R_{ED}$ at $90\%$ C.L. for three setups: P2SO, DUNE+T2HK, and DUNE+T2HK+P2SO with different conditions. }
    \label{tab:my_label}
\end{table}

\begin{figure}
    \includegraphics[scale=0.5]{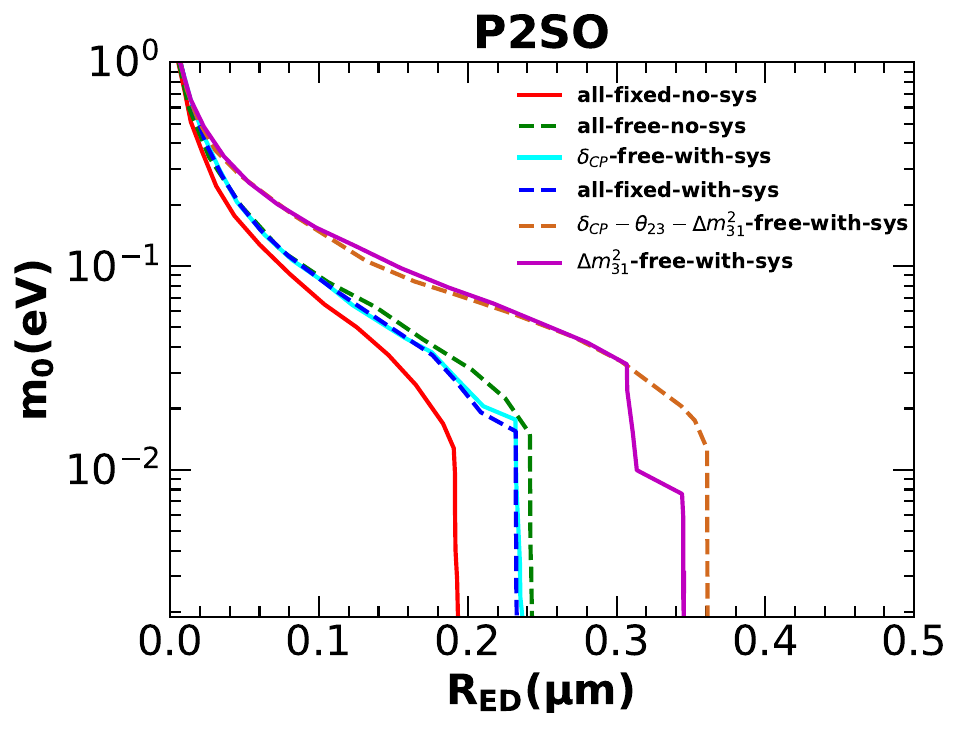}
\includegraphics[scale=0.5]{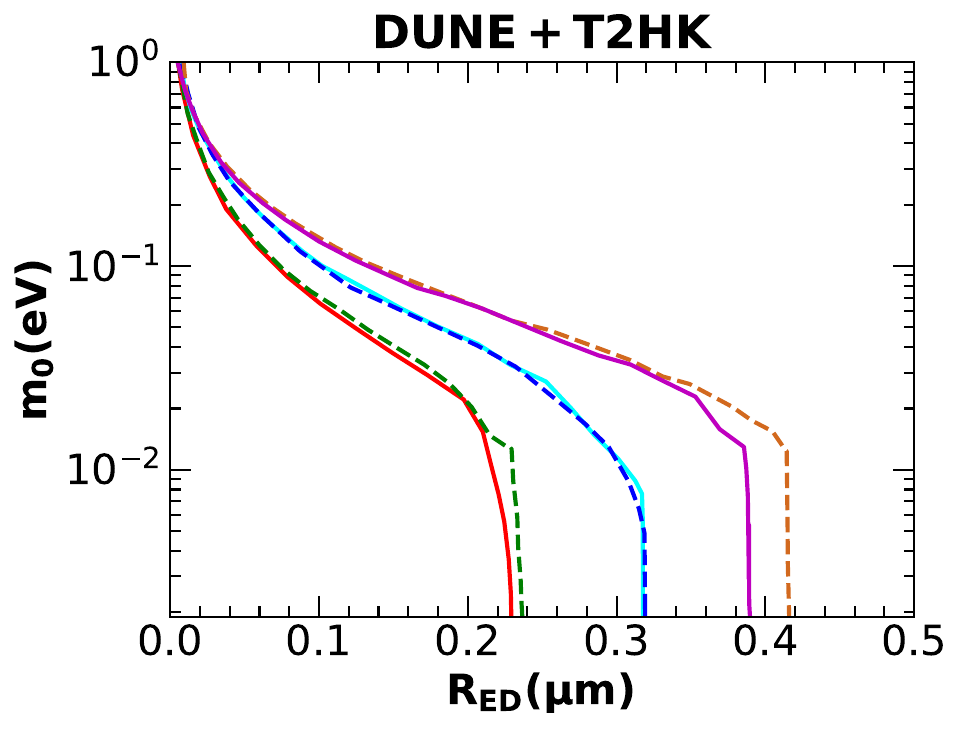}\\
\includegraphics[scale=0.5]{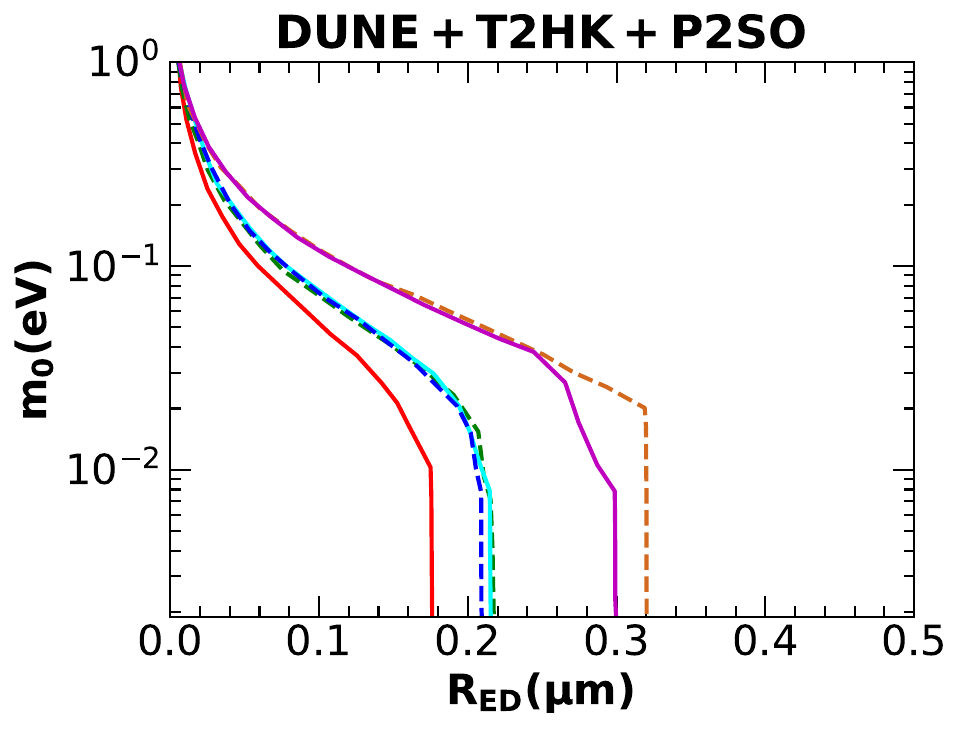}
    \caption{Bound plot in $(m_0-R_{ED})$ plane for P2SO (left of upper row), DUNE+T2HK (right of upper row), and DUNE+T2HK+P2SO (lower row). The bounds are shown for different conditions, as indicated in the legend.}
    \label{all}
\end{figure}
We present our results in Fig.~\ref{all} and summarize them in Table~\ref{tab:my_label} under different marginalization conditions. To generate each curve in Fig. \ref{all}, we assume standard interactions (\textit{i.e.,} $R_{ED} = 0~ \mu \rm m$ and $m_0$ eV) for the true spectrum, while in the test spectrum, we vary two LED parameters to obtain $90 \%$  confidence level (C.L.) bounds. The left (right) panel of the upper row in the figure shows the constraint on $m_0-R_{ED}$ plane for the P2SO (DUNE+T2HK) experiment. The lower row shows the bound plots for the combination of DUNE, T2HK, and P2SO.  In each panel, the red solid and green dashed curves represent results without any systematic errors, while the cyan solid, blue dashed, brown dashed, and purple solid curves show the results with systematic errors. To analyze the effect of each oscillation parameter, we consider different maginalization combinations. For instance, the red solid and blue dashed curves are generated when all oscillation parameters are fixed in the test spectrum of the $\chi^2$, whereas the cyan solid (brown dashed) curve is produced by varying $\delta_{\rm CP}$ ($\delta_{\rm CP}, \theta_{23}$, and $\Delta m_{31}^2$). The purple curve is obtained by allowing only $\Delta m_{31}^2$ to vary. 

From the figure we see that the weakest bound on $R_{ED}$ corresponds to $m_0 = 0$ eV and as $m_0$ increases, the bound becomes more stringent. Further, we observe that for each experimental setup, the strongest bound on $R_{ED}$ arises in the ideal case where all oscillation parameters are known and no systematic errors are included. The bound gets weaken as we include systematic errors in the analysis. We also notice significant changes in the $R_{ED}$ bound when different oscillation parameters are allowed to vary. When all the oscillation parameters are kept free, we obtain the weakest bound on $R_{ED}$. Additionally, when only $\Delta m_{31}^2$ is marginalized, the bound on $R_{ED}$ becomes much weaker compared to the marginalization of other parameters. This behavior holds true across all setups: P2SO, DUNE+T2HK, and DUNE+T2HK+P2SO. Among the different setups, the bound on $R_{ED}$ from the P2SO experiment alone is stronger than the combination of DUNE and T2HK. However, the synergy of all three experiments provides a more stringent bound on the LED parameter compared to the DUNE+T2HK and P2SO setups individually. When all the oscillation parameters are marginalized and when we include systematics, the bound on $R_{ED}$ at $90\%$ C.L. is $0.361 ~\mu \rm m$ for P2SO, $0.414 ~\mu \rm m$ for DUNE+T2HK, and $0.320 ~\mu \rm m$ for DUNE+T2HK+P2SO. If we compare the current bound with our results then we find that our results at 90\% C.L. corresponding to DUNE+T2HK+P2SO including systematics and considering all the oscillation parameters known ($R_{ED} < 0.208~\mu \rm m$ ) is better than the current bound obtained by the combined results from MINOS/MINOS+, Daya Bay, and KATRIN ($R_{ED} < 0.250~\mu \rm m$) \cite{Forero:2022skg}.

\label{led-sys}
\begin{figure}
    \centering
\includegraphics[width=0.5\linewidth]{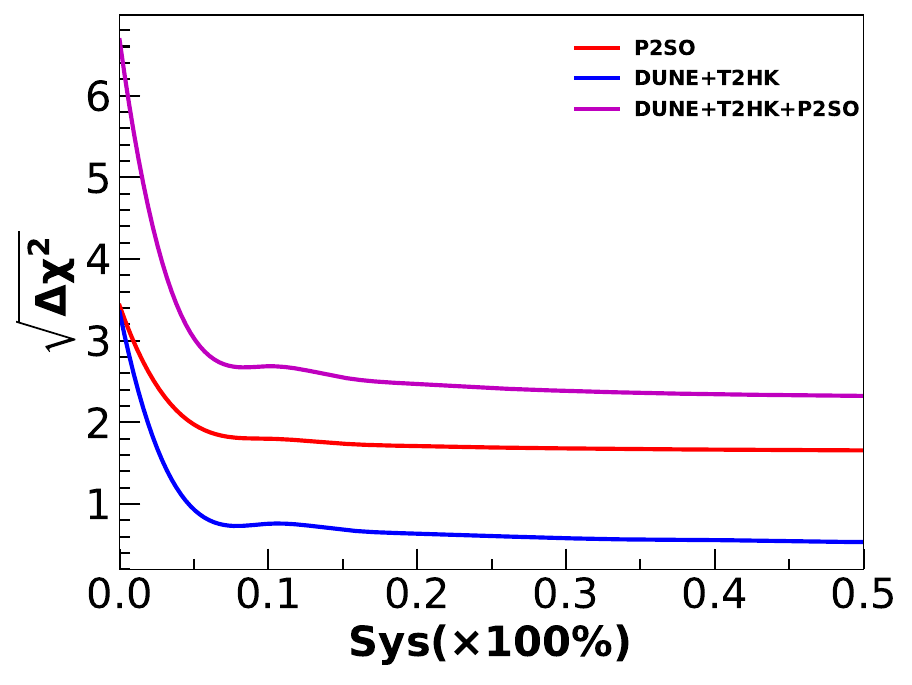}
    \caption{Bounds as a function of systematics for P2SO, DUNE+T2HK, and DUNE+T2HK+P2SO.}
    \label{sys}
\end{figure}   
To examine the impact of systematics on the bound on $R_{ED}$ into more detail, in Fig.~\ref{sys} we show the bound as a function of systematic error with the red, blue, and purple curves representing the P2SO, DUNE+T2HK, and DUNE+T2HK+P2SO setups, respectively. The systematics mentioned in the figure represent the uncertainty from the overall normalization errors.  We have generated this panel for $R_{ED} = 0.5~ \mu$m and $m_0 = 0$ eV. From this figure we see that sensitivity drops significantly as systematic uncertainty increases from 0\% to 10\%. For DUNE+T2HK+P2SO, the sensitivity falls from more than $6 \sigma$ \footnote{Here  $\sqrt{\Delta \chi^2}=\sigma$ for $1$ degree of freedom.} to less than $3 \sigma$ when systematics increases from 0\% to 10\%.  Beyond that, the sensitivity mostly remains flat.

\subsubsection{Physics sensitivities in presence of LED\\}
\label{led-sensi}

\begin{figure}
      \centering
\includegraphics[height=65mm, width=80mm]{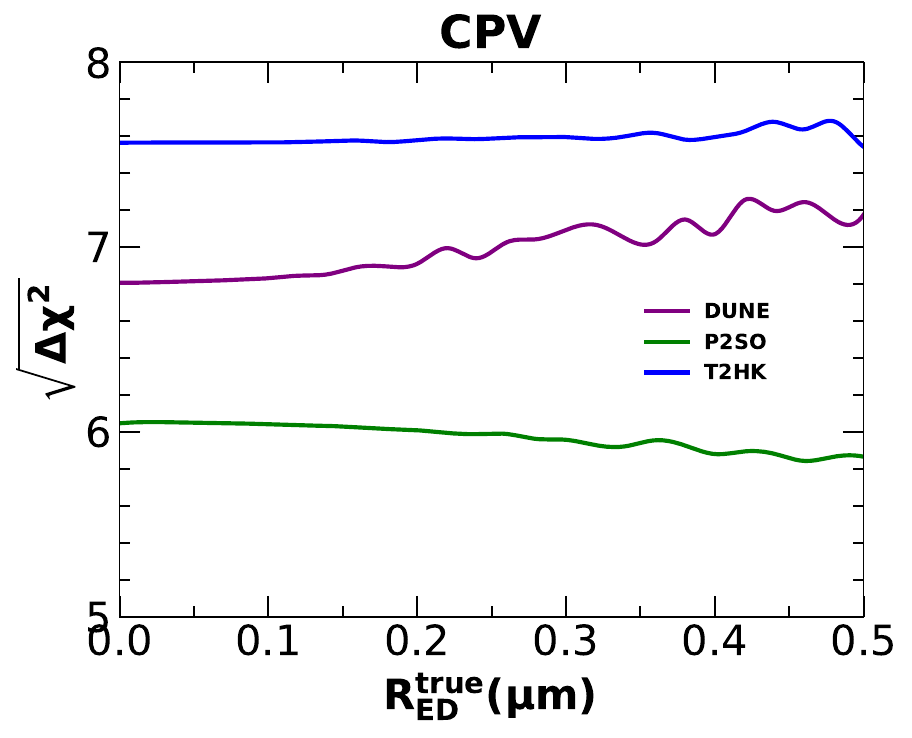}
\includegraphics[height=65mm, width=80mm]{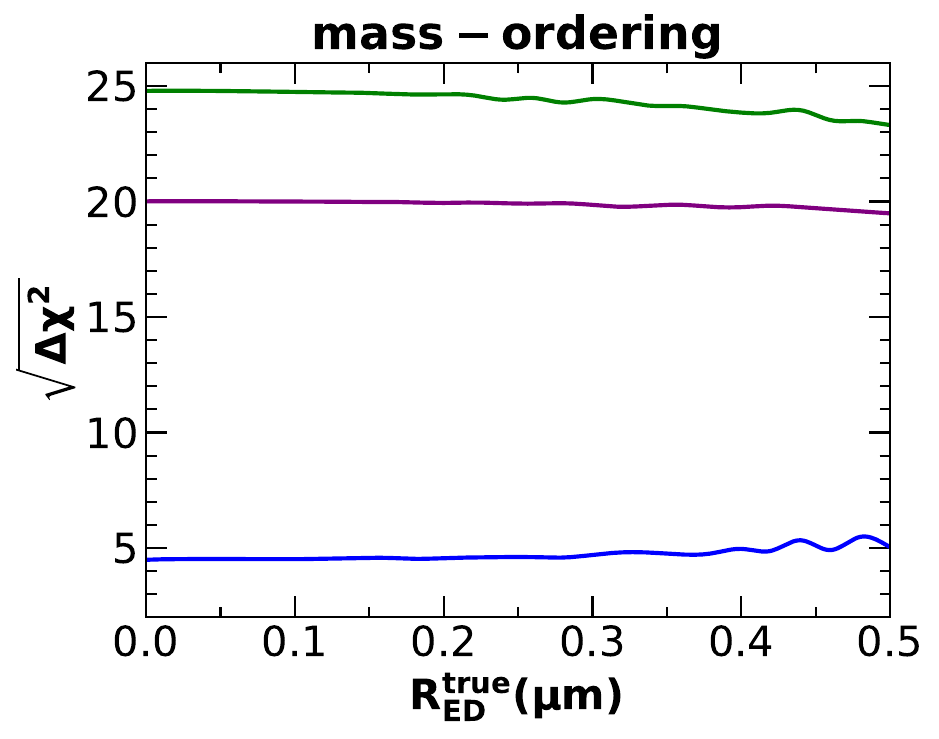}\\      \includegraphics[height=65mm, width=80mm]{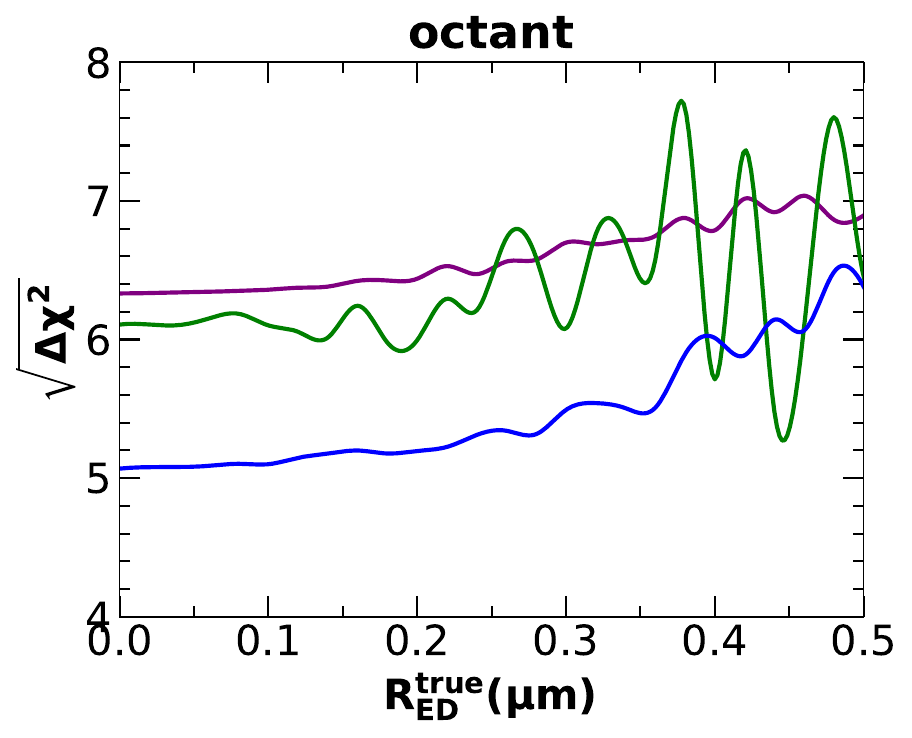}
      \caption{Left (right) panel of upper row shows the CPV (mass ordering) sensitivity as a function of $R_{ED}$ (in $\mu$m). Lower row depicts the octant sensitivity of $\theta_{23}$ as a function of $R_{ED}$. In each panel, purple, green, and blue curves are the sensitivity results for DUNE, P2SO, and T2HK experiments respectively.}
      \label{cpv}
  \end{figure}

In this subsection, we examine how LED parameters affect CP violation (CPV), mass ordering, and the octant sensitivity of the neutrino oscillation experiments under consideration. CPV sensitivity refers to the ability of an  experiment to distinguish a CP-conserving phase ($\delta_{\rm CP} = 0^{\circ} ~\rm{or} ~ 180^{\circ}$)  from a CP-violating phase. Mass ordering sensitivity is the capability of any experiment to exclude a true hierarchy from the test hierarchy. Octant sensitivity represents the capability of an experiment to distinguish the lower octant from the upper octant of the atmospheric mixing angle $\theta_{23}$. Fig.~\ref{cpv} illustrates our results where we have considered LED in both true and test spectrum of the $\chi^2$ with $m_0 = 0$ eV. For this figure, we have considered the true value of $\delta_{\rm CP}$ as $270^\circ$. In the top row, the left panel shows the CPV sensitivity as a function of the LED parameter $R_{ED}$, while the right panel shows mass ordering sensitivity with respect to $R_{ED}$, lower row depicts the octant sensitivity. For mass ordering sensitivity analysis, we assume normal ordering in the true spectrum,
while in the test scenario, we consider inverted ordering. In each panel, the purple, green, and blue curves represent results for the P2SO, DUNE, and T2HK experiments, respectively. 
\begin{table}
    \centering
    \begin{tabular}{||c||c||c||}
    \hline
    \hline
      Experiment  & Sensitivity & $|\Delta \sqrt{\Delta \chi^2}|$ \\
      \hline
      \hline
      P2SO    & CPV & $0.18 ~\sigma$\\
      \hline
        &  Mass & $1.47 ~\sigma$\\
        \hline 
          &  Octant & $0.35~\sigma$\\
          \hline 
          \hline
          DUNE & CPV & $0.37 ~\sigma$\\
          \hline
            &  Mass & $0.52 ~\sigma$\\
            \hline 
              &  Octant & $0.56 ~\sigma$\\
              \hline
              \hline
              T2HK & CPV & $0.02 ~\sigma$\\
              \hline
                &  Mass & $0.53 ~\sigma$\\
                \hline
                  &  Octant  & $1.31 ~\sigma$\\
                  \hline
                  \hline
    \end{tabular}
    \caption{Change of sensitivity when the $R_{ED}$ parameter varied from 0 to $0.5 ~\mu m$.}
    \label{ch}
\end{table}

From all the three panels of this figure we can see that the sensitivity remains almost flat when $R_{ED}$ is not very large i.e., < 0.3 $\mu$m. This signifies the fact that presence of LED does not affect the sensitivity to the standard parameters of the long-baseline neutrino oscillation experiments for small $R_{ED}$. This behavior can be understood by looking at Eq.~\ref{eq app}. From this equation we can see that the new physics coefficients involving LED, i.e., $A$, $B$ and $C$ are very small as compared to the leading order standard scenario term. Because of this, the change in the sensitivity in presence of LED will only become relevant when $R_{ED}$ is very high. This is also evident from this figure where we see the sensitivity changes slightly for $R_{ED}$ > 0.3 $\mu$m. In the lower panel, we observe a very rapid change in the sensitivity for P2SO when $R_{ED}$ > 0.3 $\mu$m. We have checked that, this wiggles appear due to the matter effect. If we consider no matter effect in P2SO, then the curve becomes smoother. 

In Table~\ref{ch} we have listed the change of the sensitivity corresponding to CPV, mass ordering and octant for all the three experiments when $R_{ED}$ varies from 0 $\mu$m to 0.5 $\mu$m.

\section{Invisible Decay}
\label{sec:decay}

This section explores the propagation of neutrinos in the context of invisible neutrino decay. Following a structure similar to the discussion of LED, we begin by outlining the theoretical framework for invisible decay of $\nu_3$ state. We then analyze its influence on oscillation probabilities and event rates. Subsequently, we estimate bounds on the decay parameter in the context of the P2SO experiment, considering the effects of minimization over different oscillation parameters.
Next, we discuss how invisible neutrino decay impacts CP violation and the octant sensitivity of P2SO. These results are further explained using analytical expressions to provide deeper insights into the numerical findings. It is important to note that this section focuses exclusively on P2SO, as studies on invisible neutrino decay for the other two experiments are already available in the literature. Additionally, combining results from multiple experiments is not addressed here, given that the projected limits from individual future experiments are expected to be significantly stronger than current constraints. Lastly, we do not explore the mass ordering in the presence of decay. This is because, in the inverted ordering, $\nu_3$ is not the heaviest mass state and thus has a lower probability of decay compared to $\nu_2$. The latter is already tightly constrained by solar neutrino data \cite{Berezhiani:1992xg, Berryman:2014qha, Huang:2018nxj} and observations from supernova SN1987A \cite{Frieman:1987as}.

\subsection{Theoretical framework}
\label{sec:decay-theo}
We consider a BSM scenario where  neutrino could decay to a lighter neutrino and a massless Majoron  at the tree level, described by the following interaction Lagrangian~\cite{Barger:1999bg, Lindner:2001fx, Beacom:2002cb}:
\begin{align}
  \mathcal{L}=\frac{1}{2}[g_s \overline{\nu_{i}}\nu_j J + ig_p \overline{\nu_{i}} \gamma_5 \nu_j J ],
\end{align}
where $g_s$ and $g_p$ are the scalar and pseudoscalar coupling constants respectively.

This framework permits the heaviest state $\nu_j$, to decay into the lighter one $\nu_i$ and a Majoron $J$. In case of invisible decay, this lighter state could be a sterile neutrino ($\nu^\prime$) and provide the $\nu_j \rightarrow \nu^\prime + J$ decay. To isolate the effects of neutrino decay, we assume alignment of the mass basis ($\nu^\prime$) and flavor basis ($\nu_s$) of the sterile neutrino.  This assumption eliminates any potential impact of sterile neutrinos on the oscillation probabilities and there exists a unitary relation that links the mass and flavor bases:
\begin{align}
    \begin{pmatrix}
    \nu_\beta\\
    \nu_s 
    \end{pmatrix} =\begin{pmatrix}
    U & 0\\
    0&1 
    \end{pmatrix}\begin{pmatrix}
    \nu_j\\
    \nu^\prime 
    \end{pmatrix} ,
\end{align}
where $U$ is the PMNS matrix, $\beta = e, \mu, \tau$ and $j=1,2,3$. In this study, we examine the decay of the $\nu_3$ state in the presence of matter, described by the following Hamiltonian:
 \begin{align}
      H_{tot} = U[H_{vac}+ H_{dec}]U^\dagger +H_{mat},
      \label{Eq:Hamilt-NuD}
\end{align}

where \begin{align}
H_{vac} =\frac{1}{2E_\nu} \begin{pmatrix}
          0~&~0&0\\
          0~&~\Delta m^2_{21}&0\\
          0~&~0&\Delta m^2_{31}
      \end{pmatrix} ,
      ~H_{dec} = \frac{\Delta m^2_{31}}{2E_\nu}\begin{pmatrix}
          0~&~0&0\\
          0~&~0&0\\
          0~&~0&- i\gamma_m
      \end{pmatrix},~ H_{mat}= \rm Diag ~(V_{CC},~0,~0).
\end{align} 
Here $H_{vac}$, $H_{mat}$, and $H_{dec}$ correspond to Hamiltonian components that govern neutrino propagation in vacuum, matter, and in the presence of decay, respectively. The term $\gamma_m = \frac{1}{\Delta m^2_{31}}\frac{m_3}{\tau_3}$ is always a real quantity. The decay term makes the Hamiltonian non-Hermitian, resulting in a loss of total probability, which indicates neutrino depletion in the system, \textit{i.e.}, $\sum\limits_{\beta = e,\mu, \tau} P_{\alpha\beta} < 1$ (where $\alpha= e,\mu, \tau$). The transition probability from flavor $\alpha$ to $\beta$ has been analytically calculated for a two-flavor oscillation scenario in Ref. \cite{Chattopadhyay:2021eba}, and for a three-flavor scenario in Refs. \cite{Lindner:2001fx, Abrahao:2015rba, Ghoshal:2020hyo, Chattopadhyay:2022ftv, Banerjee:2023sxj, Gronroos:2024jbs}.

\subsection{Oscillation probability and events in presence of decay}
\label{prob-decay}

In this section, we examine the impact of invisible decay on neutrino oscillation probabilities. The expression for the appearance and disappearance probabilities in the presence of decay can be written as \cite{Gronroos:2024jbs,Chattopadhyay:2022ftv}

\begin{equation}
    \begin{split}
        P_{\mu e} &=s_{13}^2s_{23}^2\frac{1+\gamma_m^2}{(A_m-1)^2+\gamma_m^2}\Big\{1-2\cos\big[2(A_m-1)\Delta\big]\text{e}^{-2\gamma_m\Delta}+\text{e}^{-4\gamma_m\Delta}\Big\}\\
        &+\frac{\alpha s_{13}\sin{2\theta_{12}}\sin{2\theta_{23}}}{(A_m-1)^2+\gamma_m^2}\frac{\sin(A_m\Delta)}{A_m} \Bigg\{(A_m-1-\gamma_m^2)\sin(A_m\Delta+\deltaCP)\\
        &+\sin\big[(A_m-2)\Delta-\deltaCP\big](A_m-1-\gamma_m^2)\text{e}^{-2\gamma_m\Delta} \\
        &+A_m\gamma_m\bigg[\cos(A_m\Delta+\deltaCP)-\cos\big[(A_m-2)\Delta-\deltaCP\big]\text{e}^{-2\gamma_m\Delta}\bigg]\Bigg\},
    \end{split} 
    \label{Eq:Pmue}
\end{equation}
\begin{equation}
        P_{\mu\mu} = 1-s_{23}^2\Big(1-\text{e}^{-4\gamma_m\Delta}\Big) - c_{23}^2s_{23}^2\Big[1-2\cos(2\Delta)\text{e}^{-2\gamma_m\Delta}+\text{e}^{-4\gamma_m\Delta}\Big],
        \label{Eq:Pmumu}
\end{equation}
where $s_{ij} (c_{ij})= \sin{\theta_{ij}}( \cos{\theta_{ij}})$, $\alpha = \Delta m^2_{21}/\Delta m^2_{31}$, $\Delta = \Delta m^2_{31}L/4E_{\nu}$ and $A_m = 2V_{cc} E_\nu/\Delta m^2_{31}$. 
\begin{figure}[htbp]
      \includegraphics[height=60mm, width=80mm]{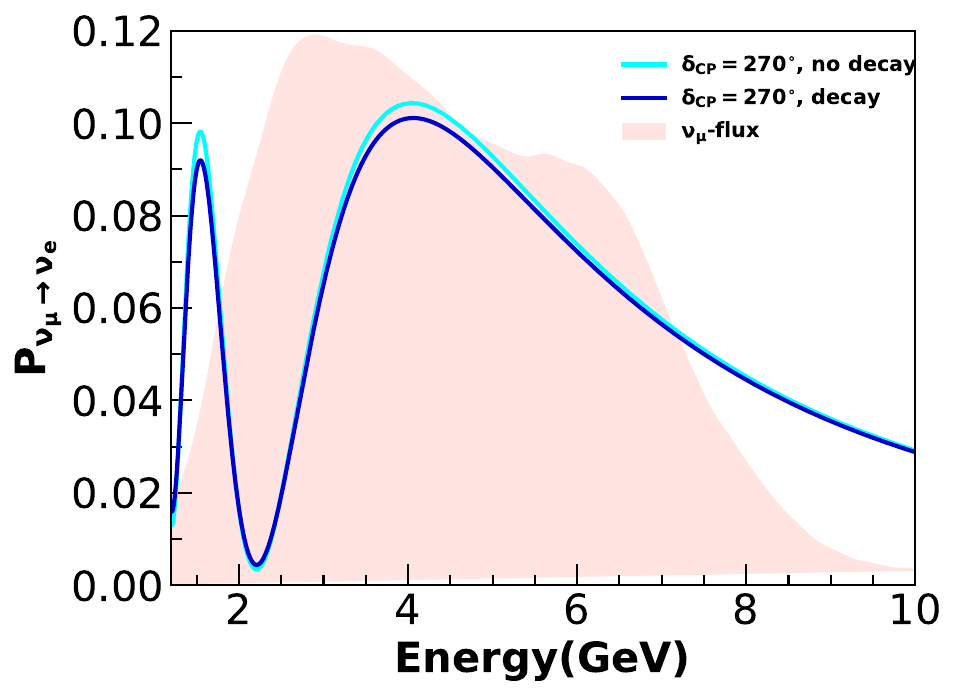}
      \includegraphics[height=60mm, width=80mm]{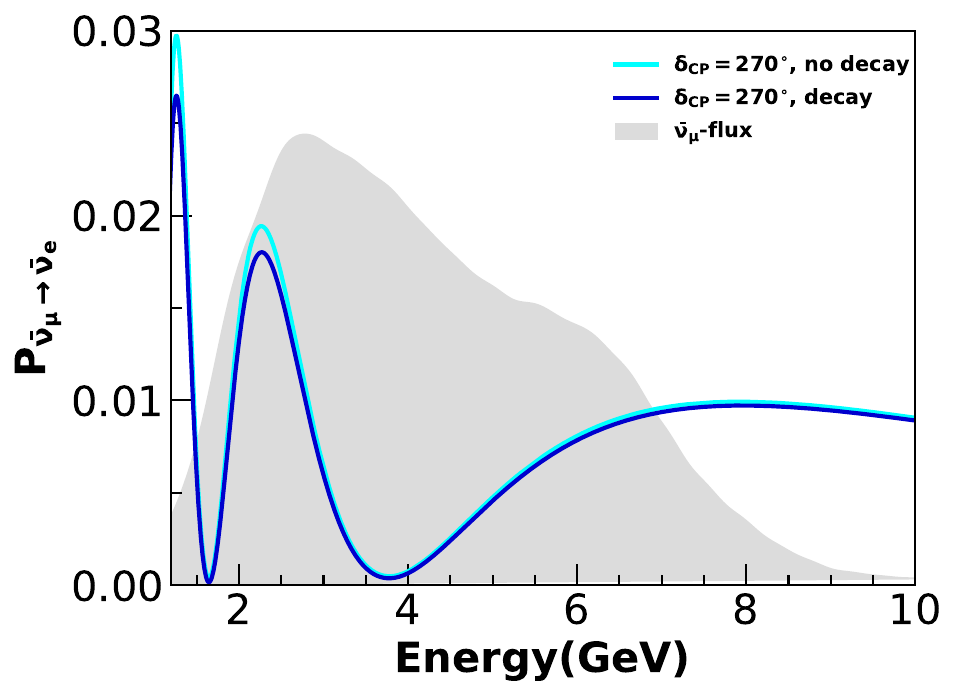}\\
      \includegraphics[height=60mm, width=80mm]{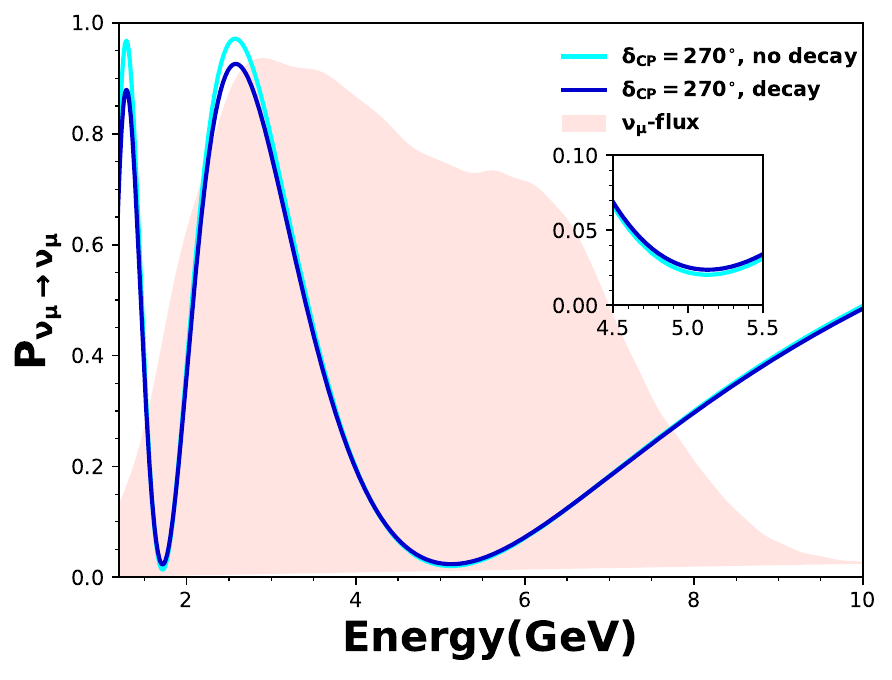}
      \includegraphics[height=60mm, width=80mm]{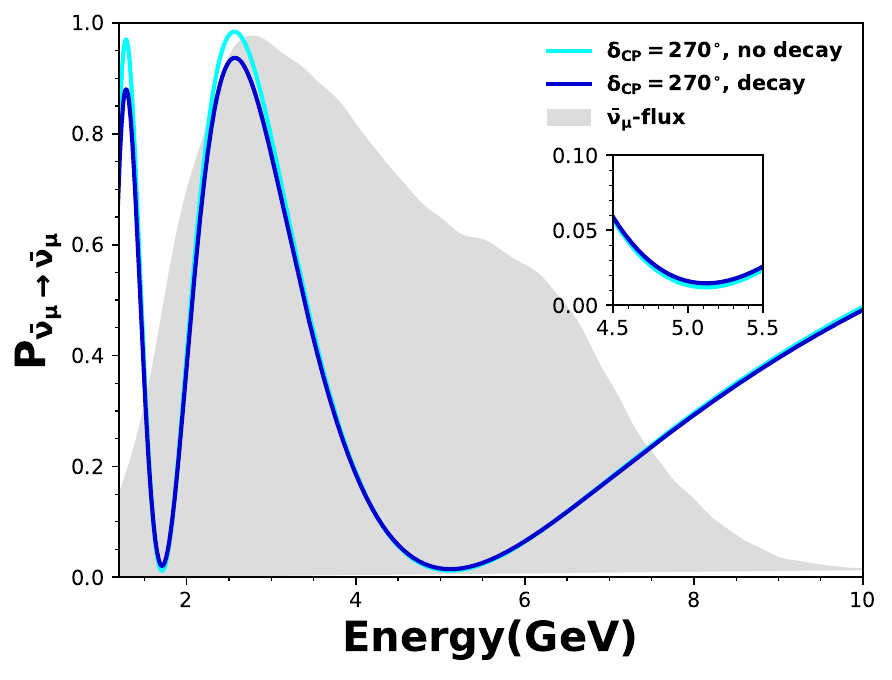}
          \caption{The appearance and disappearance probabilities for both the neutrino and antineutrino modes, considering the cases of decay and no decay, in P2SO experiment. The shaded area illustrates the flux in arbitrary unit associated with the respective oscillation channel. }
      \label{Fig:probability-NuD}
  \end{figure}
Fig. \ref{Fig:probability-NuD} displays the  appearance and disappearance probabilities for neutrino and antineutrino modes in the upper and lower panels, respectively. These probabilities are plotted for $\deltaCP = 270^\circ$. The orange and gray shaded regions depict the same fluxes as mentioned in Fig. \ref{prob3}. Throughout all panels, the cyan and blue curves represent the scenario of no decay and decay respectively. We consider a fixed value for the decay parameter $\tau_3/m_3 = 3\times 10^{-11}$ s/eV in all the decay probabilities. We observe that for both neutrinos and antineutrinos, decay leads to a decrease in the appearance probability at its peak value, a slight reduction in the disappearance probability at the point where $P_{\mu\mu}$ peaks, and a small increase at the point where $P_{\mu\mu}$ reaches its minimum. These behaviors can be explained using the probability expressions given in Eqs. \ref{Eq:Pmue} and \ref{Eq:Pmumu}. This we do in the next paragraph.

It is crucial to note that the condition $\gamma_m = 0$ represents the case of a stable neutrino. When decay is present, the two coefficients in Eq. \ref{Eq:Pmue} decrease as functions of $\gamma_m$. Furthermore, near the oscillation peak, the rate of this decrease becomes steeper due to the presence of terms that are multiplied by integer powers of $\text{e}^{-\gamma_m \Delta}$. In case of disappearance channel, we focus around 4.9 GeV energy because dip contributes more in the sensitivity analysis in the presence of decay. The bottom row of Fig. \ref{Fig:probability-NuD}, illustrates that the probability of disappearance rises with the introduction of decay at the dip.  This aspect of probability can be understood using the Eq.~\ref{Eq:Pmumu}, where $\gamma_m$ dependent terms, namely the integer powers of $e^{-\gamma_m \Delta}$, decreases with increasing $\gamma_m$. Further, near $E_\nu \sim 4.9$ GeV, $\cos 2\Delta$ approaches $-1$ and therefore, the term involving $\cos 2\Delta ~e^{-2\gamma_m \Delta}$ transforms into an increasing function, thereby influencing the overall characteristics of the disappearance probability.
\begin{figure}[htbp]
     \includegraphics[height=60mm, width=80mm]{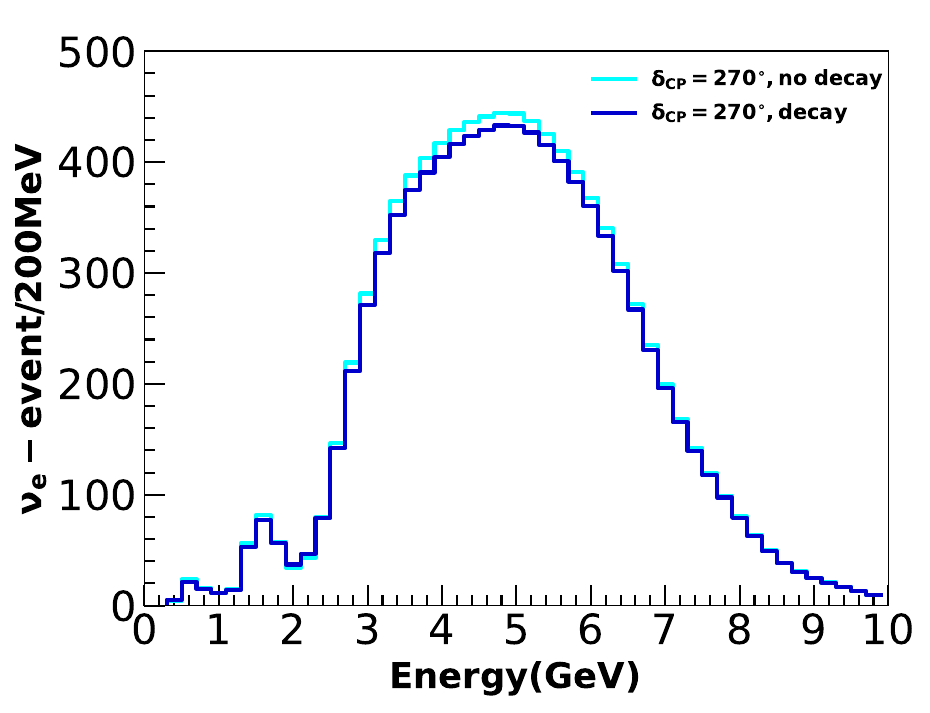}
     \includegraphics[height=60mm, width=80mm]{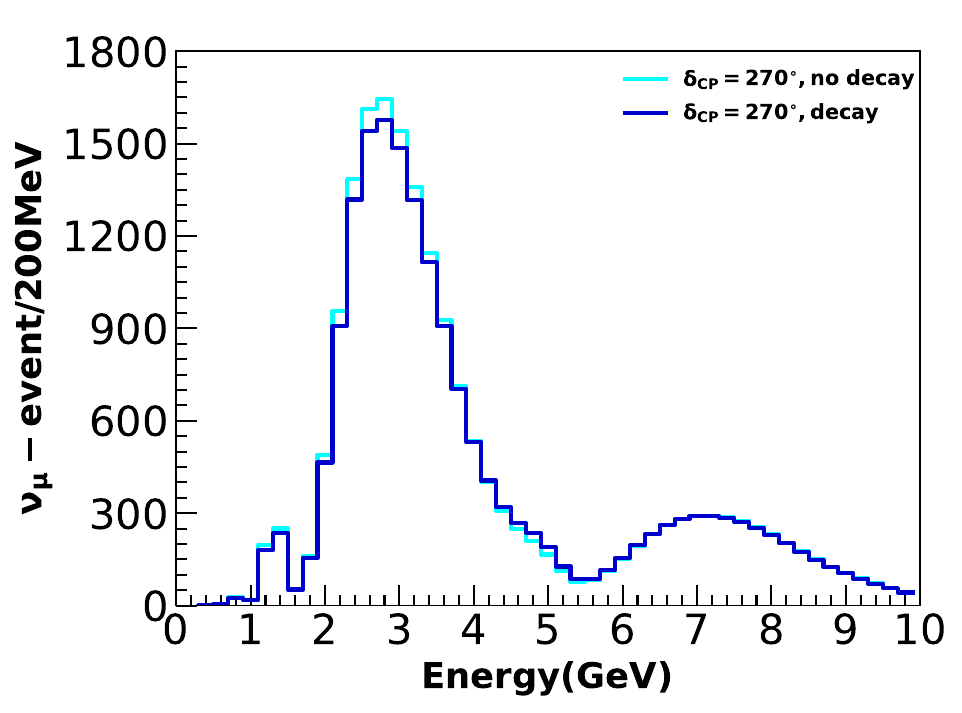}
         \caption{Event rates for $\nu_e$ appearance and $\nu_\mu$ disappearance channels for P2SO experiment with and without neutrino decay.}
     \label{fig:event-NuD}
 \end{figure}
 Fig. \ref{fig:event-NuD} provides the event rates for $\nu_e$ and $\nu_\mu$ corresponding to a specific value of $\tau_3/m_3$ and $\deltaCP = 270^\circ$ with 200 MeV energy bin. The events represented by cyan correspond to no decay scenario whereas events in blue corresponding to a decay scenario with $\tau_3/m_3 = 3 \times 10^{-11} ~\rm s/eV$. It is noted that for $\nu_e$ events, there is a decrease in event rates around the energy ($\sim$ 4.9 GeV), where events peak when compared to the no decay scenario, whereas an opposite trend is observed in the $\nu_\mu$ events. As energy increases, the impact of decay on the event count becomes negligible.

\subsection{Results}
\label{sec:results}

\subsubsection{Bound on neutrino decay\\}
\label{subsubsec:decay-bound}

\begin{figure}[h!]
     \centering
     \includegraphics[height=70mm, width=72mm]{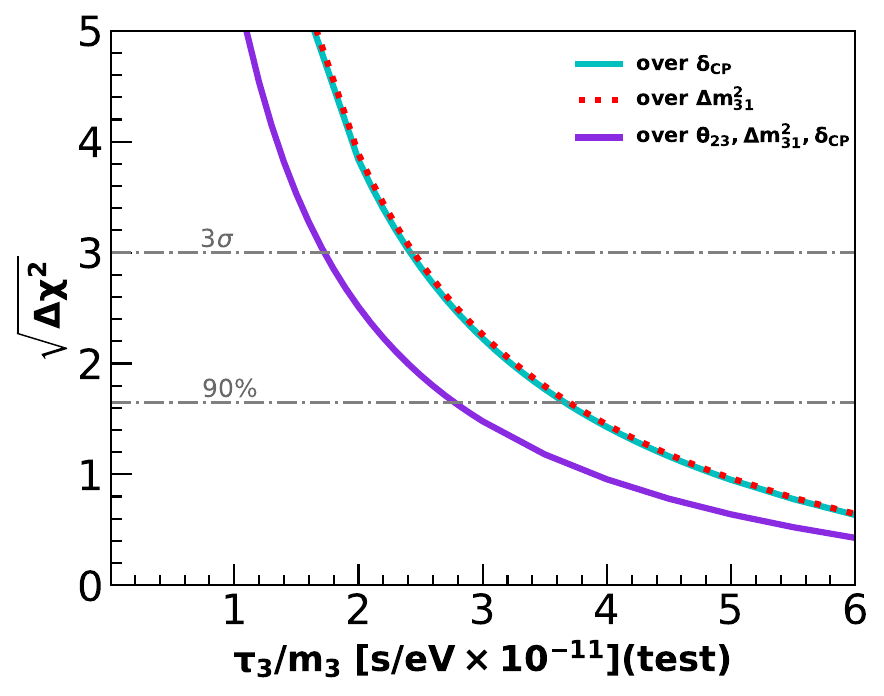}
         \caption{ Sensitivity bound plot for neutrino decay parameter with different marginalization conditions. }
     \label{Fig:bound-NuD}
 \end{figure}

The capability of the P2SO experiment to constrain the decay parameter $\tau_3/m_3$ is assessed by simulating results under the assumption of stable neutrinos in the true scenario and decaying neutrino in the test values. The results are depicted in Fig.~\ref{Fig:bound-NuD}. The curves in purple, cyan, and dotted orange represent three distinct scenarios of marginalization. The cyan curve corresponds to marginalization solely over $\deltaCP$, while the dotted orange curve is derived from marginalization exclusively over $\Delta m^2_{31}$. The purple curve results from marginalizing simultaneously over $\theta_{23}$, $\Delta m^2_{31}$, and $\deltaCP$. It is evident that marginalizing only over $\Delta m^2_{31}$ and $\deltaCP$ yields a similar effect on the bound curve; however, the inclusion of $\theta_{23}$ significantly decreases sensitivity.

For a comprehensive comparison of the $\tau_3/m_3$ parameter values obtained in this study with those from other experiments, we have included Table~\ref{Tab:NuD-limits}, which presents the 3$\sigma$ C.L. bounds from P2SO alongside MOMENT, ESSnuSB, DUNE, and T2HK experiments. For additional comparisons with other experiments,  see the Table 3 in Refs. \cite{Choubey:2020dhw, Dey:2024nzm}. We find that P2SO gives slightly better bound than MOMENT and ESSnuSB experiments but not as good as DUNE and T2HK experiments, which is because of large background of P2SO.
\begin{table}[h!]
\begin{center}
\begin{tabular}{|l||*{5}{c|}} \hline
\backslashbox{\hspace{0cm}$\tau_3/m_3$}{\hspace{0cm}Exp.} & \makebox[6.4em]{MOMENT \cite{Tang:2018rer}} & \makebox[6.2em]{ESSnuSB \cite{Choubey:2020dhw}} & \cellcolor{gray!30}\makebox[6.2em]{P2SO} & \makebox[6.2em]{T2HK \cite{Chakraborty:2020cfu}} & \makebox[6.2em]{DUNE \cite{Dey:2024nzm}} \\ \hline \hline
\hspace{0.3cm}$3\sigma$\hspace{0.3cm} C.L. [s/eV] & $1.6 \times 10^{-11}$ & $1.68 \times 10^{-11}$ & \cellcolor{gray!30} $2.11 \times 10^{-11}$ & $2.72 \times 10^{-11}$ & $4.22\times 10^{-11}$ \\ \hline
\end{tabular}
\caption{Projected sensitivities of MOMENT, ESSnuSB, P2SO, DUNE, and T2HK experiments in constraining the $\tau_3/m_3$ parameter. Shaded region indicates the bounds obtained in this work.}
\label{Tab:NuD-limits}
\end{center}
\end{table}

\subsubsection{Physics sensitivity in presence of neutrino decay}
\label{decay-sensi}
\begin{enumerate}
    \item \textbf{CPV sensitivity:}\\
In this section, we examine the impact of invisible neutrino decay on the sensitivity measurements of  CPV at P2SO experiment. The left panel of Fig.~\ref{Fig:CPV-NuD} depicts the sensitivity to CPV as a function of $\tau_3/m_3$ (true). 
The parameter $\tau_3/m_3$ is kept same in both true and test cases simultaneously. It is observed that as the $\tau_3/m_3$ value increases, the sensitivity to distinguish between the CP conserving and violating scenarios also increases, eventually stabilizing at higher values of $\tau_3/m_3$. An increased $\tau_3/m_3$ clearly approaches the scenario without decay. This implies that the sensitivity to CPV deteriorates in the presence of neutrino decay. To understand this, we show the $\nu_e$ appearance probability as a function of $\delta_{\rm CP}$ for different values of $\tau_3/m_3$ in the right panel of Fig. \ref{Fig:CPV-NuD}. The black curve represents the scenario without decay, while the red, green, and blue curves correspond to $\tau_3/m_3$ values of $5 \times 10^{-12}$ s/eV, $1.7 \times 10^{-12}$ s/eV, and $0.8 \times 10^{-12}$ s/eV, respectively. As the value of $\tau_3/m_3$ decreases from the no decay scenario, the red, blue and green curves exhibit a tendency to flatten, thereby indicating a lack of distinction between CP conserving and violating cases. This nature of curve can also be explained by the probability expression given in Eq. \ref{Eq:Pmue} \cite{Gronroos:2024jbs}. The term which is mainly contributing to the difference in probabilities at two specific values of $\deltaCP$ for different values of $\gamma_m$ is:
\begin{align}
   \frac{\alpha s_{13}\sin{2\theta_{12}}\sin{2\theta_{23}}}{(A_m-1)^2+\gamma_m^2}\frac{\sin(A_m\Delta)}{A_m} \sin\big[(A_m-2)\Delta-\deltaCP\big](A_m-1-\gamma_m^2)~\text{e}^{-2\gamma_m\Delta}.
\end{align}

 \begin{figure}
     \centering
     \includegraphics[height=69mm, width=75mm]{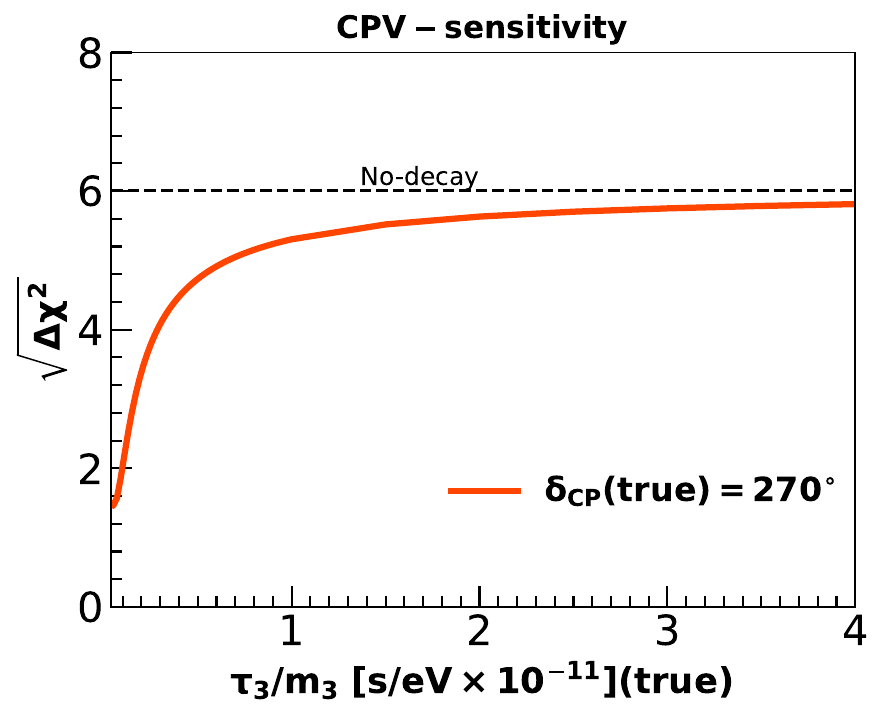}
     \includegraphics[height=65mm, width=79mm]{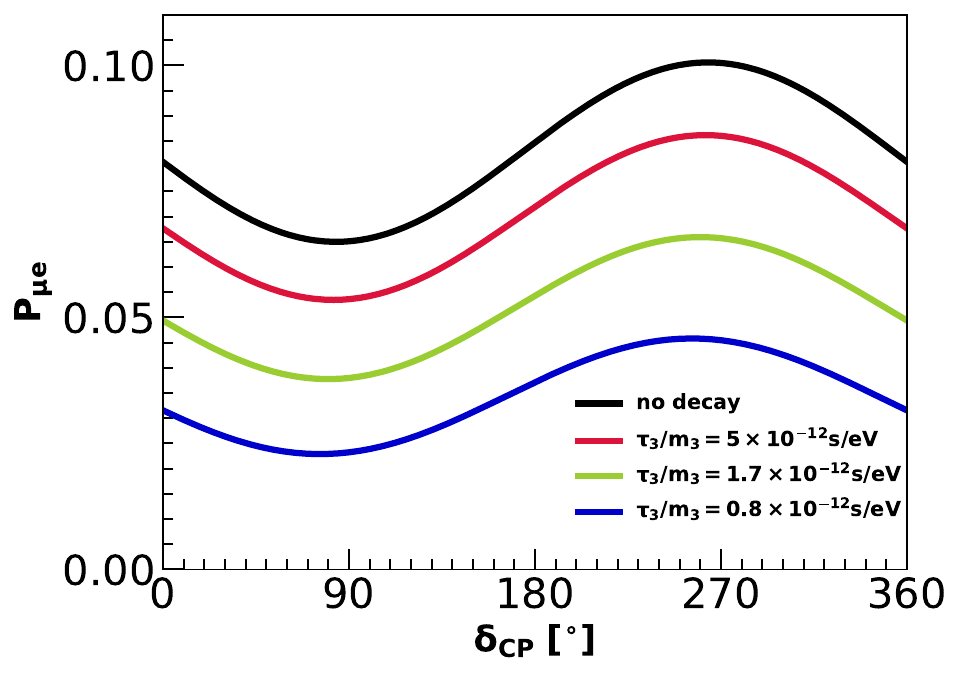}
     \caption{For P2SO experiment, the left panel shows the expected sensitivity to CP violation in the presence of neutrino decay, while the right panel illustrates the oscillation appearance probability. The curves in the right panel correspond to different values of $\tau_3/m_3$ and are plotted at a neutrino energy of 4.9 GeV.}
     \label{Fig:CPV-NuD}
 \end{figure}

The above mentioned term is positive and a decreasing function of $\gamma_m$ for $\deltaCP = 270^\circ$. We take two particular values of $\deltaCP$ in test and checked that the $\chi^2$ minima occurs at $\deltaCP = 180^\circ$.  Now, the same term is found to be negative and an increasing function of $\gamma_m$ for $\deltaCP = 180^\circ$. The  $\tau_3/m_3$ parameter shows opposite behavior compared to $\gamma_m$. Thus, a decrease in $\tau_3/m_3$ leads to a reduced separation between the CP violation and CP conservation probabilities. This is the main reason, the CP violation sensitivity decreases when one deviates from the standard scenario.

\item \textbf{Octant sensitivity}\\

The capability of the P2SO experiment to exclude the  wrong octant is shown in the top panel of Fig. \ref{Fig:Oct-NuD} which presents the combined $\sqrt{\Delta\chi^2}$ from both disappearance and appearance channels as a function of $\tau_3/m_3$ (true).  
The analysis of the plot reveals that the sensitivity begins to rise from a low value of $\tau_3/m_3$, subsequently reaching a peak before it starts to decline. Eventually, it converges with the standard scenario at larger values of $\tau_3/m_3$.
\begin{figure}
     \centering
          \includegraphics[height=70mm, width=80mm]{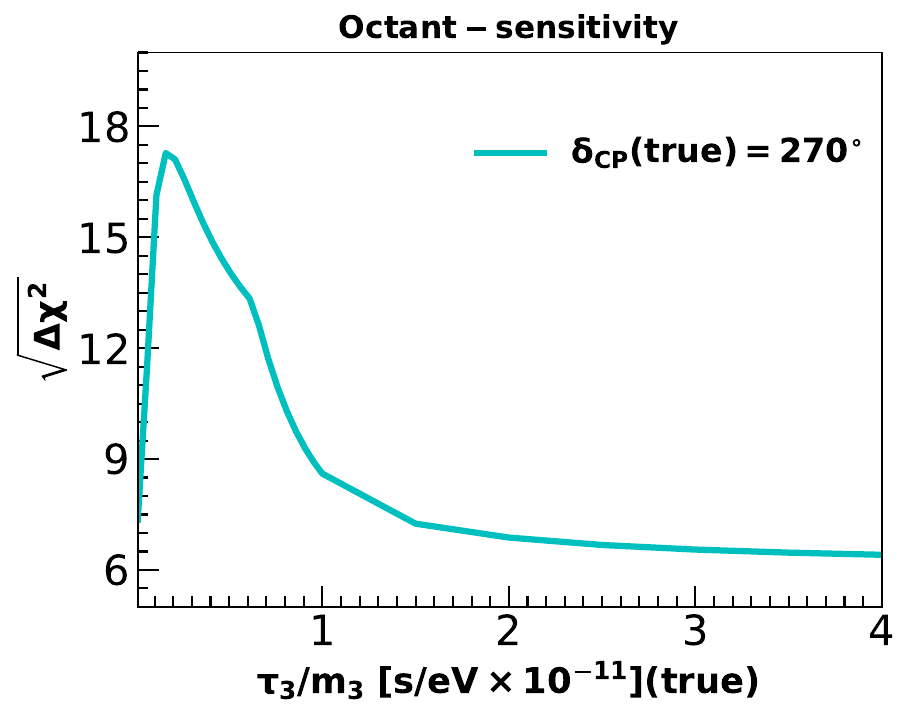}\\
      \includegraphics[height=65mm, width=80mm]{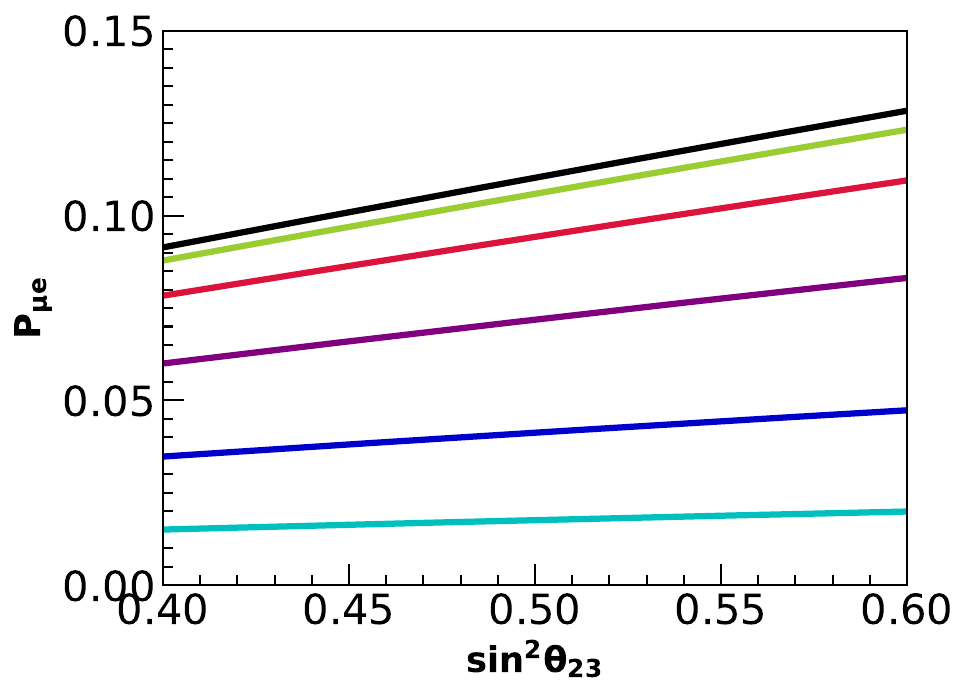}
     \includegraphics[height=65mm, width=80mm]{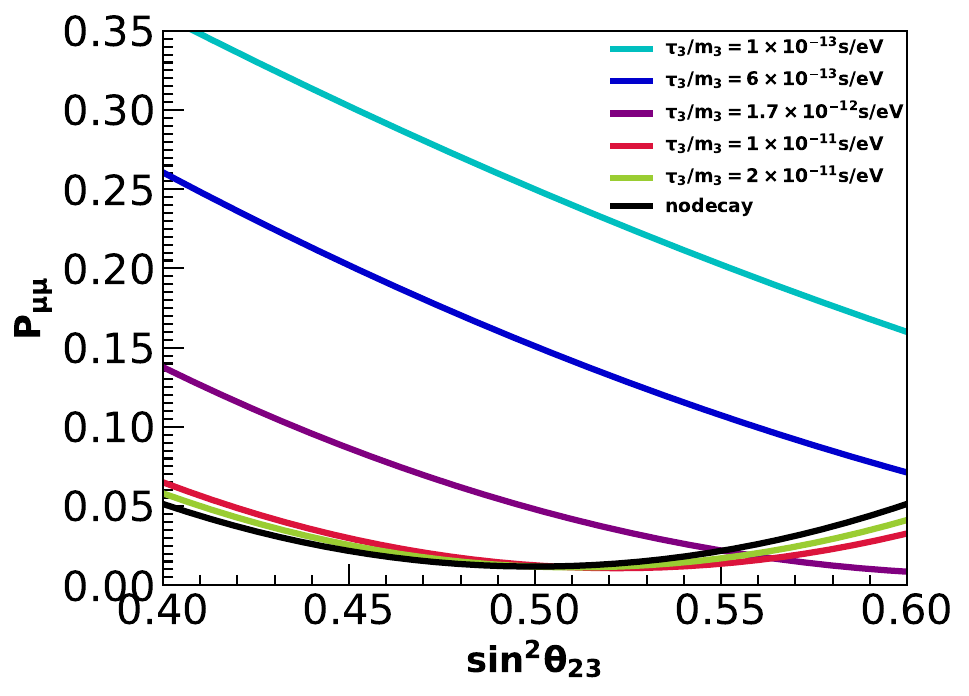}
     \caption{For P2SO experiment, in the top panel, the expected octant sensitivity is presented in the presence of neutrino decay  while the bottom left (bottom right) panel portrays oscillation appearance (disappearance) probability as a function of $\sin^2{\theta_{23}}$ for various values of $\tau_3/m_3$, including the no decay scenario in black color. The probability panels correspond to neutrino energy of 4.9 GeV and $\delta_{CP} = 270^\circ$.}
     \label{Fig:Oct-NuD}
 \end{figure}

To understand this behavior, we plot $P_{\mu e}$ and $P_{\mu \mu}$ as a function of $\sin^2{\theta_{23}}$ for various values of $\tau_3/m_3$ as shown in left and right column of lower row of Fig.~\ref{Fig:Oct-NuD}. We will first examine $P_{\mu e}$.  It is observed that as we progress from a scenario without decay to one that includes decay, the slope of curves keep decreasing, thereby indicating a lower sensitivity with decreasing $\tau_3/m_3$.  The $P_{\mu e}$ channel does not adequately explain the combined characteristics of the octant sensitivity curve. We will now turn our attention to the $P_{\mu \mu}$ curve displayed in the right panel. From this panel the octant degeneracy in the standard scenario is completely visible as one can have exactly same value of $P_{\mu \mu}$ for two different values of $\theta_{23}$; one lying in lower octant and the other lying in upper octant. However, once decay is introduced, the degeneracy gets lifted and as $\tau_3/m_3$ decreases, the disappearance channel becomes more sensitive to the octant.
This phenomenon is a result of the degeneracy between the $\tau_3/m_3$ and $\theta_{23}$ parameters, as elaborated in Ref. \cite{Chakraborty:2020cfu}. 
This pattern is further clarified from the $P_{\mu \mu}$ expression given in Eq.~\ref{Eq:Pmumu}, where the second term is sensitive to the octant. But as it contains a damping factor $\text{e}^{-4\gamma_m\Delta}$, sensitivity increases with $\gamma_m$ i.e., with decreasing $\tau_3/m_3$. However, this does not explain the peak in the octant sensitivity curve. The peak of octant sensitivity curve appears mainly because of the functional form of $\chi^2$ distribution. To understand it better, we refer to Table ~\ref{tab:s23-chisq}. In this table, for different values of $\tau_3/m_3$, we have listed the two disappearance probabilities : $P_{tr}$ at the true values of oscillation parameters, and $P_{test}$ at the test values of oscillation parameters that yield the minimum of $ \Delta \chi^2$. We also present the corresponding $\chi^2$ values in the last column. The data indicates that as we begin with the smallest value of $\tau_3/m_3$ and move towards standard no decay scenario, the difference between the true and test probabilities tends to decrease. However, for the $\chi^2$ distribution function, we see that it rises to a peak at  $\tau_3/m_3 = 1.7\times 10^{-12}$ s/eV, which signifies an increase in octant sensitivity as $\tau_3/m_3$ rises. Following this peak, it begins to decrease, ultimately approaching a minimal octant sensitivity in the standard no decay case.
 
 \begin{table}[h!]
     \centering
     \begin{tabular}{||c||c|c|c|c||}
     \hline\hline
        $\tau_3/m_3$ [s/eV]&$P_{tr}$ &$P_{test}$& $P_{tr}-P_{test}$ & $\chi^2 = \frac{(P_{tr}-P_{test})^2}{P_{tr}}$ \\ \hline \hline
          
          $\hspace{6mm}1.0 \times 10^{-13}$\hspace{6mm} & \hspace{6mm}0.3097\hspace{6mm} & \hspace{7mm}0.2552\hspace{7mm} & \hspace{6mm}0.0545\hspace{6mm} & 0.0095 \\ \hline
         $6.0 \times 10^{-13}$  & 0.2194 & 0.1655 &  0.0539 & 0.0132 \\ \hline
          $1.7 \times 10^{-12}$ & 0.1015 & 0.0578 & 0.0437 & 0.0188 \\ \hline
         $1.0 \times 10^{-11}$ & 0.0364 & 0.0137 & 0.0227 & 0.0142 \\ \hline
         $2.0 \times 10^{-11}$ & 0.0307 & 0.0191 & 0.0116 & 0.0044 \\ \hline
         \hspace{-0.4cm}no decay & 0.0253 & 0.0260 & -0.0006 & $1.54 \times 10^{-5}$ \\ \hline \hline
     \end{tabular}
     \caption{Obtained values of $\nu_\mu$-disappearance probabilities for P2SO experiment and neutrino energy of $\rm 4.9~ GeV$ at different values of $\tau_3/m_3$ for true value of $\sin^2{\theta_{23}} = 0.448$ and $\deltaCP=270^\circ$.}
     \label{tab:s23-chisq}
 \end{table}

\end{enumerate}
\section{Concluding remarks}
\label{conclu}

Future long-baseline experiments will play a pivotal role in exploring physics beyond the standard three-neutrino paradigm. In this paper, we focus on two such new physics scenarios, namely LED and neutrino decay, in the context of long-baseline neutrino experiments, with a special emphasis on the P2SO detector. The LED model offers an elegant solution to the hierarchy problem and can naturally generate small neutrino masses. We introduce three 5-dimensional fermionic fields in addition to the SM fields. Compactifying the fifth dimension on a circle of radius $R_{ED}$ produces KK states, which mix with the lowest-lying active neutrinos, thereby affecting neutrino oscillations. The impact of LED on neutrino propagation can be described by two free parameters: $m_0$ and $R_{ED}$. We illustrate the effects of these parameters on probabilities and events and provide a simplified expression of the appearance probability. The presence of LED parameters introduces a fast-changing phase which results in rapid oscillations (wiggles) in the probability and it also causes a reduction in the probabilities. We present bound on the LED parameters under various marginalization conditions for the proposed experiments: P2SO, the combination of DUNE and T2HK and the synergy of three experiments. Marginalization over the $\Delta m^2_{31}$ parameter significantly affects the sensitivity. We also examine the effect of systematic uncertainties and observe that it impacts the sensitivity significantly. In the ideal scenario (\textit{i.e.}, without any systematic error and all the oscillation parameters are measured without any uncertainty), the combined experiments (DUNE+T2HK+P2SO) can exclude $R_{ED} > 0.175 ~\mu$m at $90\%$ C.L. However, including both uncertainties weakens the bound, allowing exclusion of values of $R_{ED} > 0.320 ~\mu$m at $90\%$ C.L. Notably, P2SO provides a much stronger bound on $R_{ED}$ compared to DUNE and T2HK combined. We further investigate the impact of LED on CPV, neutrino mass ordering and octant of $\theta_{23}$, and found that the LED parameters have only a mild effect on these sensitivities if $R_{ED}$ is small. 

Next, we investigate the impact of the invisible decay of $\nu_3$ into a sterile state ($\nu'$) and a Majoron in the context of the P2SO experiment. We present the probability and event rates in the presence of the decay parameter. Neutrino decay leads to an overall decrease in probability, with a slight increase observed in the disappearance channel near the first oscillation maximum. The marginalization over $\theta_{23}$ has a significant effect on constraining the decay parameter $\tau_3/m_3$. Using the P2SO setup, one can exclude $\tau_3/m_3 < 2.11 \times 10^{-11}$ $\rm{s/eV}$ at $3\sigma $ C.L. Additionally, we examine the effect of decay on CPV and octant of $\theta_{23}$ sensitivities as functions of $\tau_3/m_3$ and found that the presence of decay reduces the CP violation sensitivity in the P2SO experiment. In contrast, octant sensitivity exhibits a unique behavior with respect to $\tau_3/m_3$. Starting with a small $\tau_3/m_3$ value, the $\chi^2$ initially increases up to a certain point, then decreases as $\tau_3/m_3$ grows. This distinctive pattern arises due to the degeneracy between $\theta_{23}$ and the $\tau_3/m_3$ parameter.

\section{Acknowledgement}

 PP and PM want to thank Prime Minister’s Research Fellows (PMRF) scheme for its financial support. SR is supported by the NPDF grant (PDF/2023/001262) from SERB,
Government of India. The work of MG has been in part funded by Ministry of Science and Education of Republic of Croatia grant No. KK.01.1.1.01.0001 and European Union under the NextGenerationEU Programme. Views and opinions expressed are however those of the author(s) only and do not necessarily reflect those of the European Union. Neither the European Union nor the granting authority can be held responsible for them. RM would like to acknowledge University of Hyderabad IoE project grant no. RC1-20-012. We gratefully acknowledge the use of CMSD HPC facility of
University of Hyderabad to carry out the computational works.

\bibliography{main}   
\end{document}